%% file: lmwt-paper.tex
\documentclass[aps,prd,twocolumn,superscriptaddress,nofootinbib]{revtex4}

\pdfoutput=1

\usepackage{amsmath}
\usepackage{graphicx}
\usepackage{bbold} 
\usepackage{xspace} 
\usepackage[caption=false]{subfig} 
\usepackage{grffile}
\usepackage{slashed}
\usepackage{rotating,multirow} 

\usepackage{simplewick} 

\usepackage[prependcaption,textsize=tiny]{todonotes}

\newcommand\one{\mathbb{1}}
\newcommand \spinhalf {spin-$\frac{1}{2}$\xspace}
\newcommand \mpcac {m_{\mathrm{PCAC}}}
\newcommand\pb{\overline{\psi}}
\newcommand{\tr}{\mathrm{T}}
\newcommand\tra{\mathrm{tr}}
\newcommand\pmp{\xi_{+}}
\newcommand\pmm{\xi_{-}}
\newcommand{\xib}{\bar{\xi}}
\newcommand\Gb{\overline{\Gamma}}
\newcommand\nubar{\overline{\nu}}
\newcommand\Sgroup[2]{\mathrm{#1}(#2)}
\newcommand\su[1]{\Sgroup{SU}{#1}}
\newcommand\so[1]{\Sgroup{SO}{#1}}
\newcommand\uone{\Sgroup{U}{1}}

\newcommand{\ch}{\chi}
\newcommand{\ad}{{\dot{\alpha}}}
\newcommand{\al}{\alpha}

\newcommand{\be}{\beta}
\newcommand{\mc}{\zeta}
\newcommand{\mcb}{\bar{\zeta}}

\newcommand\Nx[1]{N_{\textnormal{#1}}}
\newcommand\Nf{\Nx{f}}
\newcommand\Nc{\Nx{c}}

\begin{document}


\title{The infrared regime of $\su{2}$ with one adjoint Dirac flavour}


\author{Andreas Athenodorou}
\email{athenodorou.andreas@ucy.ac.cy}
\affiliation{Department of Physics, Swansea University, Singleton Park, Swansea SA2 8PP, UK}
\affiliation{Department of Physics, University of Cyprus, POB 20537, 1678 Nicosia, Cyprus}

\author{Ed Bennett}
\email{e.j.bennett@swan.ac.uk}
\affiliation{Department of Physics, Swansea University, Singleton Park, Swansea SA2 8PP, UK}

\author{Georg Bergner}
\email{g.bergner@uni-muenster.de}
\affiliation{Universit\"at Bern, Institut f\"ur Theoretische Physik, 
Sidlerstr.~5, CH-3012 Bern, Switzerland}

\author{Biagio Lucini}
\email{B.Lucini@swan.ac.uk}
\affiliation{Department of Physics, Swansea University, Singleton Park, Swansea SA2 8PP, UK}


\date{\today}

\begin{abstract}
$\su{2}$ gauge theory with one Dirac flavour in the adjoint
representation is investigated on a lattice. Initial results for the
gluonic and mesonic spectrum, static potential from Wilson and
Polyakov loops, and the anomalous dimension of the fermionic
condensate from the Dirac mode number are presented. The results found
are not consistent with conventional confining behaviour, instead
tentatively pointing towards a theory lying within or very near the
onset of the conformal window, with the anomalous dimension of the
fermionic condensate in the range $0.9 \lesssim \gamma_* \lesssim
0.95$. The implications of our work for building a viable theory of
strongly interacting dynamics beyond the standard model are discussed.  
\end{abstract}


\maketitle


\section{Introduction}
\label{sect:introduction}
Even after the recent experimental identification of the Higgs
particle~\cite{Aad:2012tfa,Chatrchyan:2012ufa}, the existence of a new
fundamental interaction of which the Higgs sector is the low energy
manifestation is still an open problem. Among the proposed
possibilities, novel strong
dynamics~\cite{Weinberg:1975gm,Susskind:1978ms,Eichten:1979ah,Holdom:1981rm,Holdom:1984sk} 
is still a good candidate for a possible fundamental mechanism of
electroweak symmetry breaking. This new strong interaction is able to 
explain the observed electroweak symmetry breaking phenomenology if
the following three conditions are met: (1)
the theory must be near the onset of the conformal window; (2)
the anomalous dimension of the chiral condensate must be of
order one; and (3) a parametrically light scalar (the would-be Higgs
boson) must be in the spectrum. The first two
conditions~\cite{Yamawaki:1985zg,Appelquist:1986an} are needed for
compatibility with electroweak precision data~\cite{Peskin:1991sw}, while the third
condition is determined by the direct observation of the Higgs boson
and no other previously unknown nearby state. Until very
recently, even the possible existence of a strongly interacting
quantum field theory for which any of those conditions arose was unclear. In the
last few years, much progress has been achieved on these theoretical
questions, thanks to a combination of methods and techniques. A
crucial role has been played by numerical investigations using lattice
techniques, which---among other results---have pinned down an example 
of a gauge theory in the conformal window, namely $\su{2}$ gauge theory
with two adjoint Dirac
flavours~\cite{DelDebbio:2009fd,Hietanen:2009az,Catterall:2009sb,Bursa:2009we,DelDebbio:2010hu,DelDebbio:2010hx,Bursa:2011ru,Catterall:2011zf,DeGrand:2011qd}
(see~\cite{Catterall:2007yx,DelDebbio:2008zf,Hietanen:2008mr,Catterall:2008qk} for earlier simulations 
of the model). Although conformal strong dynamics can still explain 
electroweak symmetry breaking, an anomalous dimension around one is
required~\cite{Chivukula:2010tn}. This condition rules out a possible phenomenological role
played by $\su{2}$ with two adjoint Dirac flavours in its simplest version: in fact, the most
recent measurements of the anomalous dimension $\gamma^{\star}$ for
this model give
$\gamma^{\star} = 0.38(2)$~\cite{DelDebbio:2013hha}, which is well below the acceptable value.
Other candidate near-conformal theories exist, but with the possible
exception of SU(3) with eight flavours~\cite{Aoki:2014oha}, whose
near-conformal nature needs to be better explored, all of them have
an anomalous dimension that is too small
(see~\cite{Giedt:2012it,Kuti:2013ku} for recent overviews of lattice calculations).

At this stage, it is a fundamental problem to understand whether large
anomalous dimensions can arise in the context of conformal or
near-conformal gauge theories. Although the anomalous dimension is small
at the perturbative zeros of the beta function, large anomalous
dimensions might arise near or at the lower end of the conformal
window. As mentioned above, for SU(2) gauge theories with adjoint Dirac
fermions, lattice studies show that the model with two flavours is
infrared conformal with a small anomalous dimension. Hence, a remaining
potential way to observe a large anomalous dimension is to consider the
case of a single Dirac flavour, or equivalently two Majorana (or Weyl) 
fermions. 

Analytically, this theory could be seen as the large scalar
mass limit of ${\cal N} = 2$ super Yang--Mills with gauge group
$\su{2}$, with supersymmetry completely
broken by a non-zero mass term for the scalar. Despite the
existence of an interpolating parameter (in this specific case, the mass
of the scalar) confinement in ${\cal N} =2$ super Yang--Mills theory
does not trivially imply confinement in the limit in which the scalar
decouples. In ${\cal N} = 2$ super Yang--Mills, confinement is
known to arise through the dual superconductor
mechanism resulting from magnetic monopole
condensation~\cite{Seiberg:1994rs}. However, this mechanism does
not immediately generalise to the non-supersymmetric case, since
non-trivial effects (e.g.\ the fate of the monopoles when decoupling the
scalar) enter the interpolating theory. Exploratory lattice studies
exist for SU($N$) gauge theory with a single adjoint Dirac flavour in the
large-$N$ limit~\cite{Hietanen:2012ma,Bringoltz:2012sy,Gonzalez-Arroyo:2013dva}.
These works, which exploit large-$N$ volume reduction, do not give yet
a clear picture of the infrared behaviour of the theory.  
In addition, although perturbatively the $N$-dependence of the
$\beta$-function of theories with adjoint fermions is
mild~\cite{Dietrich:2006cm}, in principle the results of those studies
might not translate immediately to the model with two colors.   

In this work, we present a first-principles investigation of the model
using numerical Monte Carlo studies of the theory discretised on a
spacetime lattice. The central result of our work is that the infrared
regime of the system is compatible with a conformal or near-conformal
behaviour, but not with a conventional QCD-like scenario in which chiral
symmetry breaking takes place. Furthermore, the
anomalous dimension (measured with two independent methods) turns out
to be 0.925(25). 

The rest of the paper is organised as follow. In
Sect.~\ref{sect:model} we present the model and the setup of our numerical
investigations. Numerical results will then be presented in
Sect.~\ref{sect:results}. Sect.~\ref{sect:discussion} discusses the implications of our investigation and
possible directions of future studies. Finally, a summary will be
presented in Sec.~\ref{sect:conclusions}. Some preliminary results
were already presented in~\cite{Athenodorou:2013eaa}. 

\section{The model}
\label{sect:model}
\newcommand\Lx[1]{\Lambda_{\textnormal{#1}}}
We consider an $\su2$ gauge theory with a single Dirac flavour in the adjoint
representation with mass $m$. Eventually, we would be interested in
understanding the properties of the theory in the massless
limit; however, numerical simulations require a non-zero fermion
mass. Hence, we deform the theory with a small fermion mass, and
study how the system approaches the massless limit. 
We stress from the outset that regardless of the phase of the theory
at zero fermion mass, with  a finite mass term chiral symmetry is
always broken, since the mass is a relevant direction for the
renormalization group trajectory.  The expectations are that if
chiral symmetry is broken in the massless limit, the response of the model to a small
varying mass will be described by chiral perturbation theory, while if
the theory is conformal the data will be in accord with the
predictions derived from a mass-deformed conformal gauge theory. A
third possibility is that the system is in the confined phase, but
close to the onset of the conformal window. In this case, it will show
mass-deformed conformal behaviour in an intermediate energy regime
between a chiral symmetry breaking scale $\Lx{IR}$ and the
ultraviolet perturbative scale $\Lx{UV}$, while chiral perturbation
theory will correctly describe the theory for energies below
$\Lx{IR}$. The latter possibility, referred commonly as {\em
  near-conformality} or {\em walking behaviour}, would be phenomenologically
  interesting, since theories near the onset of the conformal window are relevant
for gaining an understanding of strongly interacting dynamics beyond the
standard model as the mechanism of electroweak symmetry breaking.

In the following subsections, we describe the field content of the
theory, the chiral symmetry breaking pattern, and the resulting spectrum.

\subsection{Field content}
\label{sect:model:action}
In Minkowskian space, the Lagrangian of the system is given by 
\begin{eqnarray}
\label{eq:actcont}
{\cal L} = \overline{\psi}(x) \left( i \slashed{D} - m \right) \psi(x)
- \frac{1}{2} \mathrm{Tr} \left( G_{\mu \nu}(x) G^{\mu \nu} (x) \right)\ ,
\end{eqnarray}
where $\slashed{D} = \left(\partial_{\mu} + i g A_{\mu}(x) \right)
\gamma^{\mu}$, $\gamma_{\mu}$ are the Dirac matrices, $A_{\mu}(x) =
\sum_a T^a A_{\mu}^{a}(x)$ with $a = 1,2,3$, and the $T_a$ are the
generators of $\su{2}$ in the adjoint representation (i.e.\ the generators
of SO(3)). $G_{\mu \nu} = \partial _{\mu} A_{\nu}(x) - \partial
_{\nu} A_{\mu}(x) + i g [A_{\mu}(x), A_{\nu}(x)]$, with $g$ the
gauge coupling of the theory, is the field tensor, and the trace is taken
over the gauge group. Our notations for the Dirac algebra matrices and
derived symmetry operators are reported in Appendix~\ref{app:a}.

Since the theory contains a single (Dirac) flavour, described by the
spinor $\psi(x)$, at first sight the flavour structure
of~Eq.~(\ref{eq:actcont}) would seem trivial. However, since the
adjoint representation is real, it does not mix the real and
the imaginary part of the Dirac spinor. More explicitly, if $C$ is the
Dirac matrix implementing charge conjugation, we can
decompose the Dirac spinor in Majorana components
\begin{eqnarray}
\pmp = \frac{\psi + C \overline{\psi}^\tr}{\sqrt{2}} \ ,
\qquad
\pmm = \frac{\psi - C \overline{\psi}^\tr}{\sqrt{2}i} \ ,
\end{eqnarray}
such that
\begin{eqnarray}
\psi =\frac{1}{\sqrt{2}}( \pmp + i \pmm) \ ,
\end{eqnarray}
with both $\pmp$ and $\pmm$ being invariant under charge
conjugation symmetry by construction.  Eq.~(\ref{eq:actcont}) can now be rewritten as
\begin{eqnarray}
\label{eq:actcont2}
{\cal L} =\frac{1}{2} \sum_k \xib_k(x) \left( i \slashed{D} - m \right) \xi_k(x)
- \frac{1}{2} \mathrm{Tr} \left( G_{\mu \nu}(x) G^{\mu \nu} (x)
\right) ,
\end{eqnarray}
where $k = +, -$. This flavour structure (in terms of the Majorana
components) gives rise to a  non-trivial chiral symmetry breaking pattern.

\subsection{Chiral symmetry breaking pattern}
\label{sect:model:chisb}
In the notation usually applied in considerations of supersymmetry, the 
fermion part of the Lagrangian is written as
\begin{eqnarray}
&& \sum_k 
\left[
(\mcb_k)_{\ad}
(\bar{\sigma}^\mu)^{\ad \be }D_\mu(\mc_k)_\be
+
\frac{m}{2}((\mc_k)^\al (\mc_k)_\al+ h.c.)
\right]\nonumber\\
&=& \sum_k 
\left[
\mc_k^\dag
\bar{\sigma}^\mu D_\mu\mc_k
+
\frac{m}{2}(\mc_k^\tr \epsilon^\tr \mc_k + h.c.)
\right]\;,
\end{eqnarray}
with $\alpha, \beta = 1,2$ spin indices\footnote{We have made use of
  the dotted-undotted notation commonly found in supersymmetry, for
  which we refer to the specialised literature.}, Majorana spinors $\mc_k$ in the Weyl representation
\begin{equation}
  \xi_k \
 = \left(\begin{array}{c}\mc \\ \epsilon \mc^\ast \end{array}\right)_k
 = \left(\begin{array}{c}\mc_\al \\ \mcb^\ad \end{array}\right)_k\;,
\end{equation}
and $\epsilon=i\sigma_2$. The lower component of the Majorana spinor can be derived from the 
upper one using the above expression. Therefore we can ignore the lower components and form a 
$4 \Nf$ component vector out of the two upper Weyl components of each Majorana flavour, for $\Nf=1$ we get
\begin{equation}
\eta= \left(\begin{array}{c} \mc_1 \\ \mc_2 \end{array}\right)\, .
\end{equation}

In the zero mass limit the action has a $\uone_A\otimes \su{2 \Nf}$
symmetry. The $\su{2\Nf}$ part rotates the upper and lower two
components of $\eta$ into each other. The $\uone_A$ part is broken by the anomaly down to a
discrete $Z_{2\Nc}$. The remaining $Z_{2\Nc}\otimes \su{2\Nf}$ is
subject  to a spontaneous symmetry breaking if there is a non-zero
expectation value of the fermion condensate.
The condensate and the fermion mass term are invariant under the
subgroup $Z_2\otimes \so{2 \Nf}$. This is the remaining exact symmetry
group if there is a spontaneous symmetry breaking. Therefore, the
chiral symmetry breaking pattern is 
\begin{equation}
	\su{2 \Nf} \mapsto \so{2 \Nf} \ .
\end{equation}
Hence, if chiral symmetry is spontaneously broken, there are two Goldstone
bosons in this model, corresponding to the two generators
of the broken part of the symmetry. In the present case of $\Nf=1$ the complete flavour symmetry $\su{2}$ has generators $\sigma_i/2$. 
In order to mark the difference with the $\sigma_i$ acting on the two indices of the Weyl spinor, we call 
the generators in flavour space $\tau_i=\sigma_i$. The unbroken $\so{2}$ is generated by $\tau_2/2=\sigma_2/2$ 
and is equivalent to $\uone$. 

We want to arrive at a diagonal representation of the unbroken symmetry. Therefore we apply the following unitary transformation on $\eta$
\begin{align}
 \chi
  &=\frac{1}{\sqrt{2}}\left(\begin{array}{cc} 1&i\\i&1\end{array}\right)\eta
  =\frac{1}{\sqrt{2}}(1+i\tau_1)\eta
  =\left(\begin{array}{c} \ch_1 \\ \ch_2 \end{array}\right)\\
 &=P_{\textnormal{L}} \psi+P_{\textnormal{R}}(-iC)\psi^\ast=\left(\begin{array}{c} \psi_{\textnormal{L}} \\ -\sigma_2 \psi_{\textnormal{R}}^\ast \end{array}\right)\, .
\end{align}
The $P_{\textnormal{R}}$ and $P_{\textnormal{L}}$ are the projectors on the left handed ($\psi_{\textnormal{L}}$) and right handed ($\psi_{\textnormal{R}}$) part of the Dirac spinor.
The advantage of this transformation is that the unbroken generator is
now the diagonal $\tau_3/2$ and the unbroken $\so{2}$ subgroup can be rewritten as
\begin{equation}
  \label{eq:unbrokenuone}
	U = \cos \alpha + i\tau_3 \sin \alpha = e^{i\alpha\tau_3} \;.
\end{equation}
In the Majorana formulation, there is no $\uone$ symmetry for each of
the two Majorana flavours and hence one would naturally relate the unbroken
symmetry to the isospin in QCD. In the Dirac notation the unbroken
symmetry is, however, the $\uone_V$ of charge conservation. Therefore
in the following we refer to it as baryon symmetry.  Hence, in addition to parity, the
baryon charge related to the unbroken part of the chiral symmetry can
be used to classify the spectrum of the theory in the broken phase.
At this point, it is worth stressing again that chiral symmetry breaking is
expected to arise as a soft breaking at finite fermion mass,
independently of the phase of the massless theory.  

While the residual symmetry is diagonal, in this basis parity
is expressed in terms of a 
combination of charge conjugation ($\sigma_2$) and flavour rotation 
($\tau_2$). In fact, the action of parity in the original
basis  
\begin{align}
	\psi(t,\vec{x}) &\mapsto \gamma_0 \psi(t,-\vec{x}) \;, \\
	\pb(t,\vec{x}) &\mapsto \pb(t,-\vec{x})\gamma_0\;,
        \label{eq:parity}
\end{align}
determines the transformations
\begin{align}
	\ch(t,\vec{x}) &\mapsto i \sigma_2 \tau_2 \ch^\ast(t,-\vec{x})\; , \\
	\ch^\dagger(t,\vec{x}) &\mapsto -\ch^\tr(t,-\vec{x}) i\sigma_2 \tau_2\; .
\end{align}
In order to clarify the notation, we derive explicitly the
chiral symmetry breaking pattern directly in this basis. 
The chiral symmetry group $\su{2}$ commutes with the
parity transformation, since for $U \in \su{2}$
\begin{equation}
	U(i \tau_2) =(i \tau_2)U^*\; .
\end{equation}

The chiral condensate represented in the different bases of flavour and Dirac space has the following form
\begin{align}
	\pb\psi &= \frac{1}{2}\sum_k (\mc_k^\tr \epsilon^\tr \mc_k+h.c.)\\
                &= \frac{1}{2}(\eta^\tr \epsilon^\tr \eta+h.c.)\\
                &= \psi_{\textnormal{L}}^\dagger \psi_{\textnormal{R}} + \psi_{\textnormal{R}}^\dagger \psi_{\textnormal{L}} \\
                &= \ch_1^\dag \sigma_2\ch_2^\ast+\ch_2^\tr\sigma\ch_1 \\
		&= \frac{1}{2}\left(\chi^\tr \tau_1\sigma_2 \chi +
                  \chi^\dagger \tau_1\sigma_2 \chi^* \right)
                \;. \label{eq:weyl-chiral}  
\end{align}
From the last line, one can see that when the degrees of freedom are
chosen to be $\chi$ and $\chi^{\dagger}$, the chiral condensate is
left invariant under the subgroup of $U$ matrices that satisfy 
\begin{equation}
	U^\tr \tau_1 U = \tau_1 \ ,
\end{equation}
which is the $\so{2}$ subgroup generated by $\tau_3$,
Eq.~(\ref{eq:unbrokenuone}). 

Note that the symmetry breaking pattern shown here is used in
\cite{Munster:2014cja} to derive a partially quenched chiral
perturbation theory for supersymmetric Yang--Mills theory\footnote{We
  remark that~\cite{Munster:2014cja} uses a different convention on parity. See
  Appendix~\ref{app:b} for a brief discussion of the two conventions.}. The
analysis relies the small mass of the Goldstone bosons compared to the
other states in the theory. This needs to be confronted with our
results: in a conformal or near-conformal scenario the theory develops
no relevant intrinsic mass scale and the expected hierarchy of masses
is lost.   

\subsection{The spectrum}
\label{sect:model:spectrum}
In order to understand the phase of the theory from the point of view
of the chiral symmetry, we focus our attention on bilinear fermionic
operators, which can be seen as creation and annihilation operators of
physical states that play a crucial role in establishing the chiral
properties of the system.

The bilinear fermionic operators considered in this study are shown in
Tab.~\ref{tab:wdm-results}. For convenience, they are represented in
  the Dirac notation (``Dirac bilinears'') and in the
  Majorana notation (``Majorana bilinears''). For the latter case, we
  introduce the naming convention
\begin{equation}
O_{lk} (\Gamma) = \xib_{l} \Gamma \xi_{k} \ ,
\end{equation}
where the $\xi$ are the two Majorana flavours (labeled by $k,l = +,
-$) and $\Gamma$ is a Dirac matrix or a product of Dirac
matrices. Each of the Majorana and the Dirac form has some advantage:
the Dirac representation allows us to identify easily the spin quantum
numbers (reported in column ``Spin'') and the parity, while the
Majorana notation exposes the flavour structure and bridges with the
terminology often used in supersymmetry. Straightforward algebra
enables one to obtain the expression in one notation given
the expression in the other. For the sake of simplicity, we have
omitted the Weyl notation, which is particularly suited for the
$SU(2)$ quantum numbers. The Weyl notation can be easily obtained from
the Dirac notation. For instance, one finds that
  \begin{equation}
    \frac12(\psi^\tr C\gamma_5\gamma_0 \psi + \psi^\dag C
    \gamma_5\gamma_0\psi^*) = \chi^\dag \tau_2 \chi \ ,
    \end{equation}
i.e.\ the Weyl bilinear $\chi^\dag \tau_2 \chi$, transforming under the
$3^+$ representation of $\su2^{P}$, is half the sum of the Dirac
bilinears $\psi^\tr C\gamma_5\gamma_0 \psi$ and $\psi^\dag C
    \gamma_5\gamma_0\psi^*$, both in the irreduciple representation
    $3^+$ of the original flavour group, but carrying baryon charge
    $+2$ and $-2$ respectively.

\begin{sidewaystable}
  \begin{center}
		\renewcommand{\arraystretch}{1.35}
		\begin{tabular}{||c||c|c|c|c||c|c||}
                  \hline
			Spin & $\su{2}^P$ & Dirac bilinears & Majorana bilinears & $\uone^P$ & Name & Correlators \\
		\hline
		\hline
			\multirow{9}{*}{(pseudo)scalars} & $1^-$ & $\bar{\psi} \gamma_0 \gamma_5 \psi$ & $O_{++}(\gamma_0\gamma_5)+O_{--}(\gamma_0\gamma_5)$ & \multirow{2}{*}{$0^-$} & \multirow{2}{*}{pseudoscalar meson} & \multirow{2}{*}{singlet $\gamma_5$, $\gamma_0 \gamma_5$} \\
		\cline{2-4}
			& \multirow{3}{*}{$3^-$} & $\bar{\psi} \gamma_5 \psi$ & $O_{++}(\gamma_5)+O_{--}(\gamma_5)$ & & & \\
		\cline{3-7}
			&  & $\psi^\tr C \psi$ & $O_{++}(1)-O_{--}(1) + 2iO_{+-}(1)$ & $2^-$ & \multirow{2}{*}{pseudoscalar (anti)baryon} & \multirow{2}{*}{triplet $1$} \\
		\cline{3-5}
			& & $\psi^\dag C  \psi^*$ &
                        $O_{++}(1)-O_{--}(1) - 2iO_{+-}(1)$  & $-2^-$
                        & & \\
		\cline{2-7}
			& \multirow{5}{*}{$3^+$} & $\bar{\psi} \psi$; $\bar{\psi} \gamma_0 \psi$ & $O_{++}(1)+O_{--}(1)$; $O_{+-}(\gamma_0)$ & $0^+$ & scalar meson & singlet $1$, $\gamma_0$ \\
		\cline{3-7}
			& & $\psi^\tr C \gamma_5 \psi$; & $O_{++}(\gamma_5)-O_{--}(\gamma_5) + 2iO_{+-}(\gamma_5)$;  & \multirow{2}{*}{$2^+$} & \multirow{4}{*}{scalar (anti)baryon} & \multirow{4}{*}{triplet $\gamma_5$, $\gamma_0 \gamma_5$}  \\
			& & $\psi^\tr C \gamma_5 \gamma_0 \psi$ & $O_{++}(\gamma_5\gamma_0)-O_{--}(\gamma_5\gamma_0) + 2iO_{+-}(\gamma_5\gamma_0)$ & & & \\
		\cline{3-5}
			& & $\psi^\dag C \gamma_5 \psi^*$; & $O_{++}(\gamma_5)-O_{--}(\gamma_5) - 2iO_{+-}(\gamma_5)$; & \multirow{2}{*}{$-2^+$} & & \\
			& & $\psi^\dag C \gamma_5 \gamma_0 \psi^*$ & $O_{++}(\gamma_5\gamma_0)-O_{--}(\gamma_5\gamma_0) - 2iO_{+-}(\gamma_5\gamma_0)$  & & & \\
		\hline
		\hline
			\multirow{5}{*}{(axial) vectors} & $1^+$ & $\bar{\psi} \gamma_5 \vec{\gamma} \psi$; $ \bar{\psi} \gamma_0 \gamma_5 \vec{\gamma} \psi$ &  $O_{++}(\gamma_5 \vec{\gamma}) + O_{--} (\gamma_5\vec{\gamma})$; $O_{+-}(\gamma_0\gamma_5\vec{\gamma})$ & $0^+$ & axial vector meson & singlet $\gamma_5\vec{\gamma}$, $\gamma_0 \gamma_5\vec{\gamma}$ \\
		\cline{2-7}
			& $1^-$ & $ \bar{\psi} \gamma_0 \vec{\gamma} \psi$ &  $O_{+-}(\gamma_0\vec{\gamma})$ & \multirow{2}{*}{$0^-$} & \multirow{2}{*}{vector meson} & \multirow{2}{*}{singlet $\vec{\gamma}$, $\gamma_0 \vec{\gamma}$} \\
		\cline{2-4}
			& \multirow{3}{*}{$3^-$} & $\bar{\psi} \vec{\gamma} \psi$ & $O_{+-}(\vec{\gamma})$ & & & \\
		\cline{3-7}
			& & $\psi^\tr C \gamma_5 \vec{\gamma} \psi$ & $O_{++}(\gamma_5\vec{\gamma})-O_{--}(\gamma_5\vec{\gamma}) + 2iO_{+-}(\gamma_5\vec{\gamma})$  & $2^-$ & \multirow{2}{*}{vector (anti)baryon} & \multirow{2}{*}{triplet $\gamma_5\vec{\gamma}$} \\
\cline{3-5}
			& & $\psi^\dag C \gamma_5 \vec{\gamma} \psi^*$ & $O_{++}(\gamma_5\vec{\gamma})-O_{--}(\gamma_5\vec{\gamma}) - 2iO_{+-}(\gamma_5\vec{\gamma})$ & $-2^-$ & & \\
		\hline
		\end{tabular}
	\null
	\caption{Dirac and Majorana bilinears classified according to
          their $\su2^P$ and $\uone^P$ quantum numbers, together with the
          correlators used for the calculations of the corresponding masses. \label{tab:wdm-results}}
	\end{center}
  
\end{sidewaystable}

As usual, masses are extracted by looking at the large-distance
exponential decay of correlators between operators with the same
quantum numbers. For fermionic bound states, we are interested
in the $\uone^P$ quantum numbers. When expressing the relevant correlators in the original
Dirac notation, we use the conventional language of Lattice QCD. 
In particular, the words singlet and triplet do not refer to a
QCD-like isospin symmetry, which is not defined in this theory. Here
they stand for whether fermion disconnected diagrams need to be
evaluated (singlet case) or not (as it happens for the triplet). These
contributions might appear in different cases than what one expects
from QCD. For instance, to obtain a pseudoscalar meson in our theory
it is, as in one flavour QCD, unavoidable to compute disconnected
contributions. The terminology is further discussed in 
Appendix~\ref{app:c}.  The correlators with the naming convention
inherited from QCD that are needed to compute masses in a given
channel in terms of the single Dirac flavour are indicated in the last
column of Tab.~\ref{tab:wdm-results}. The naming of the states (column 
 ``Name''), which will be used as a handy reference in the following,
 is instead derived from the $\uone$ quantum numbers, which
 characterise the physical states. In particular, $2$ indicates the
baryon with charge $q = 2$, $-2$ the antibaryon with charge $q = -2$ and zero the 
scalar/vector meson (or pseudoscalar/pseudovector meson, if the parity
is negative)\footnote{The states that are called {\em
baryons} in this work are more often referred to as {\em
diquarks} in the literature of studies of gauge theories based
on a (pseudo-)real gauge group (e.g.\ $\su2$, G(2)). The same
nomenclature is sometimes used for a real representation of a
gauge group (e.g.\ the adjoint representation). This naming
convention is discussed also in Appendix~\ref{app:b}. We note
that also the spin 1/2 state introduced below has a non-trivial baryon
charge, and hence is classified as a baryon. However, with the term
{\em (anti-)baryon} we indicate in this work only the states with $q =
2$ ($q = -2$).}. 

As mentioned above, the $\su2$ quantum numbers can be easily read in
the Weyl basis. We have indicated with $1$ the 
singlet and with $3$ the triplet of the $\su2$ flavour group.
If chiral symmetry is broken, the Goldstone bosons are the
charged baryons that belong to the positive-parity triplet of the
original flavour group (quantum numbers $3^+$). Their $\uone^{P}$
quantum numbers are $\pm2^{+}$. In the Dirac notation, operators
carrying the wanted quantum numbers are $\psi^\tr C\gamma_5\gamma_0
\psi$ and $\psi^\dag C  \gamma_5\gamma_0\psi^*$. Hence, correlators
of those operators are going to play a central role: if they
identify parametrically light particles as the Lagrangian fermion mass is
sent to zero, there will be a clear support for QCD-like chiral
symmetry breaking, otherwise we will get an indication that the
theory may be in a less familiar phase or regime.

It is worth remarking that when it comes to the
definition of physical states, each choice of the fermion
notation has advantages and disadvantages. In particular, in the
Majorana notation in general states with a well-defined baryon charge
can be obtained only by combining correlators of different
bilinears. Calculations of correlators can be carried out using
elementary properties of the Dirac algebra and will not be discussed
any further. For some explicit examples, we refer
to~\cite{Athenodorou:2013eaa}.      

In addition to purely fermionic operators, one can consider gluonic
operators and mixed gluonic-fermionic operators. Calculations
involving those operators do not present any relevant difference with
respect to similar calculations performed earlier and reported in the
literature, to which we refer for further technical details (see
e.g.~\cite{DelDebbio:2009fd,DelDebbio:2010hx,Bergner:2013nwa}). Since
the physical states contributing to a correlator in a particular
channel are selected solely by their quantum numbers, in general the
large distance exponential decay of correlators with the same quantum
numbers is dominated by the same mass, which is the mass corresponding
to the ground state in that channel\footnote{However, there are
  cases in which either kinematics or dynamics prevents some states from
  appearing in certain correlators. A remarkable example in this
  category is large-$N$ QCD, for which, for instance, meson correlators do not get
  contributions from glueballs and viceversa.}. 
This is particularly relevant for the scalar, which should
emerge both in a calculation involving purely fermionic operators and
in a calculation involving purely gluonic operators with quantum
numbers\footnote{We have omitted charge conjugation, which for gauge
  group $\su2$ is always positive.} $J^{P} =0^{+}$. This channel is
particularly important for phenomenology, as in models of strongly
interacting dynamics beyond the standard model it is identified with the
Higgs boson of the standard model itself. For a novel strongly interacting
theory to be compatible with the latest experimental findings, the
scalar must turn out to be lighter than the other particles. One of
the central results of our calculation is a sufficiently precise
measurement of the notoriously noisy scalar channel that enables us to
assess with enough accuracy what is the mass difference between the
scalar and the nearest particle in the spectrum, as we shall see in
the following section.

\section{Results}
\label{sect:results}
The action of the discretised model used in our numerical study is
given by
\begin{eqnarray}
S=S_{\mathrm{G}} + S_{\mathrm{F}}
\end{eqnarray}
where
\begin{eqnarray}
  S_{\mathrm{G}} = \beta \sum_{p} {\rm Tr} \left[ 1 - U(p)  \right] 
\end{eqnarray}
and
\begin{eqnarray}
S_{\mathrm{F}} =  \sum_{x,y} {\overline \psi}(x) D(x,y) \psi(y),
\end{eqnarray}
are respectively the pure gauge part and the fermionic
contribution. Here $U(p)$ is the lattice plaquette and 
\begin{align}
  D(x,y) = \delta_{x,y} 
  - \kappa &\left[ \left(1-\gamma_{\mu}  \right) U_{\mu} (x) \delta_{y,x+\mu} \right.\\
    \nonumber 
    &+  \left. \left(1+\gamma_{\mu}  \right) U^{\dagger}_{\mu} (x-\mu) \delta_{y,x-\mu}   \right] 
\end{align}
is the massive Dirac operator. $\kappa = 1/(8+2 a m)$ is the hopping parameter,
$a$ the lattice spacing and $m$ the bare fermion mass. For further details about the lattice model, we refer
to~\cite{DelDebbio:2009fd,DelDebbio:2010hx,DelDebbio:2010hu}, where
the $\Nf = 2$ adjoint fermion case is studied for gauge group $\su2$, using similar
notations.  The simulations were done using the
HiRep code~\cite{DelDebbio:2008zf}. The Monte Carlo trajectories
used for sampling observables were generated using a Rational Hybrid
Monte Carlo (RHMC) algorithm~\cite{Clark:2006fx}.
The algebra relating the various fermion formalisms carries over to the
lattice in a straightforward way. 

Correlators among operators can be computed on a
spacetime lattice after Euclidean rotation. In particular, if $\vec{x}$ and $t$ are
respectively the spatial and temporal components of the position
vector $x$, for a bilinear $\Psi(x)$ we have
\begin{eqnarray}
\lim_{t \to \infty} \sum_{\vec{x}} \langle
\overline{\Psi}(\vec{x}_0,t_0) \Psi(\vec{x},t_0 + t) \rangle \propto
e^{- m_{\alpha} t} \ ,
\end{eqnarray}
where $m_{\alpha}$ is the lowest mass with the quantum numbers
$\alpha$ carried by $\Psi$ and the sum is over the whole spatial
volume. 

In our study, we make use of Wilson fermions, for which chiral
symmetry is explicitly broken. As a consequence, the fermion mass
gets an additive renormalisation term that shifts the chiral point
away from zero bare mass. A mass that is only subject to
multiplicative renormalisation (and hence is zero
at the chiral point) can be defined through the \emph{partially
  conserved axial current (PCAC)}. Using the Dirac notation, the PCAC
mass is defined as the large time limit of  
\begin{equation}
        a\mpcac(t)=\frac{\sum_{\vec{x}}\left\langle \partial_{0}A_{0}(\vec{x},t)P(\vec{0},0)\right\rangle }{2\sum_{\vec{x}}\left\langle P(\vec{x},t)P(\vec{0},0)\right\rangle }\;,
\end{equation}
where
\begin{align}
        A_{0}(\vec{x},t) &
        =\overline{\psi}(\vec{x},t)\gamma_{0}\gamma_{5}\psi(\vec{x},t)\\
        P(\vec{x},t) & =\overline{\psi}(\vec{x},t)\gamma_{5}\psi(\vec{x},t)\;,
\end{align}
and the time derivative is discretised using the backward-forward
symmetrised lattice difference operator (which is defined as
the difference between the values of a function in two neighbour points
divided by the lattice spacing). The lattice
technology used to define correlators and the $\mpcac$
  mass and to compute them on the lattice is by now standard (see
  e.g.~\cite{DelDebbio:2007wk} for a more extended treatment).

In our investigation of fermion correlators, we used the
$\Nf = 1$ Dirac and the Majorana formalism (see e.g.~\cite{Montvay:1995ea,Donini:1997hh} for technical
details on lattice computations involving Majorana fermions),  in some
cases performing the analysis in both ways to cross-validate 
the result. The analysis code in the Dirac formalism, used for
connected contributions to correlators, is based on HiRep, while a
code developed for studies of Super Yang--Mills
theories~\cite{Bergner:2013nwa} has been used for cross-validation of
results for triplet contributions, for calculations involving singlets and
for spin--gauge composite states. Gluonic observables (and in
particular glueball states) have been studied using the techniques
exposed in~\cite{Lucini:2010nv}. 
Our numerical results are reported below.

\subsection{Phase diagram}
\begin{figure}
	\includegraphics[width=\columnwidth]{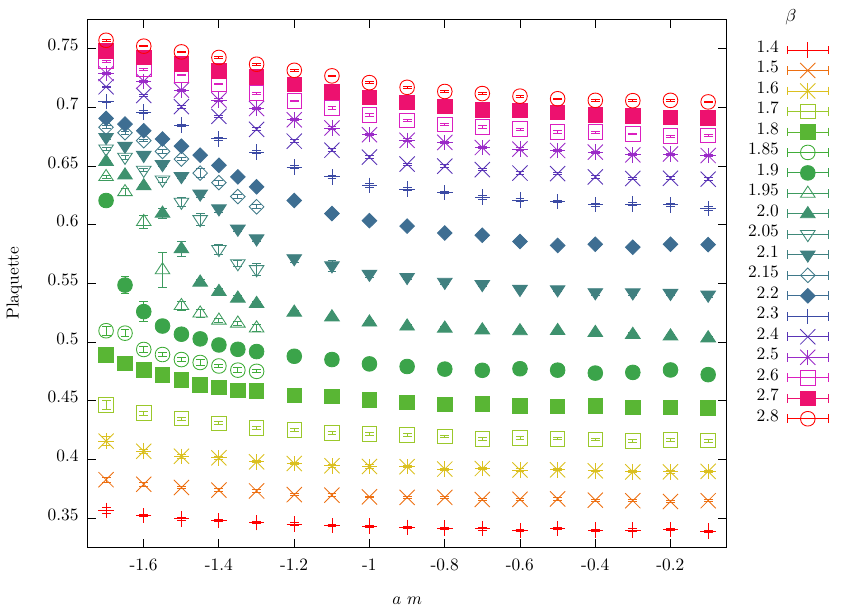}

	\vspace{-8pt}
	\caption{(Colour online) The phase diagram of the theory, showing the average plaquette on a $4^4$ lattice at $1.4\leq \beta \le 2.8$, $-1.7 \le am \le -0.1$.\label{fig:phase-diagram}}
\end{figure}
The lattice discretisation of the model 
is a good description of the latter in the limit where the inverse gauge
coupling goes to infinity. The opposite (strong coupling) limit is
generally separated from the continuum (below referred to as the
``physical region'') by a phase transition. The strong coupling phase,
dominated by lattice artefacts, is called the {\em bulk
  phase}. An order parameter for the transition from the bulk phase to
the phase connected to the continuum theory is the plaquette. Simulations
aiming at studying continuum physics need to 
make sure that the parameters are chosen in such a way that the model
is in the region connected to the continuum. A simple scan on a small
lattice allows us to perform a sensible choice of the parameters.

In the absence of a prior investigation of this theory on the lattice, a study
of the lattice phase diagram was necessary to identify the physical
region. The average plaquette was considered on a $4^4$ lattice, in
the ranges $1.4 \le \beta \le 2.8$, $-1.7 \le a m \le -0.1$, in steps of
$0.1$. Once the region of the bulk phase transition was identified,
points were added in its neighbourhood to increase the resolution to
$0.05$. The results, shown in Fig.~\ref{fig:phase-diagram}, indicate a
bulk phase transition at $\beta \approx 1.9$, $am \approx -1.65$. 

\input{tables.tex}

In order to simulate near the continuum, a large $\beta$ would be
ideally needed. However, the larger the $\beta$ the smaller the
lattice spacing. Hence, to obtain a lattice of a physically meaningful
size a large number of sites in each direction is needed. In
practical terms, this will make the simulation computationally very
costly, with the cost increasing exponentially with $\beta$. Likewise, ideally the mass should
be as small as possible. However, at fixed $\beta$ and lattice size,
strong lattice artefacts and finite size effects appear when the mass
is reduced towards the chiral limit. Hence, the minimum mass that can
be simulated depends on 
$\beta$ and on the lattice volume $V$. When choosing simulation
parameters, a compromise between the
ideally suited situation and the emergence of practical difficulties
needs to be reached, verifying {\em a posteriori} that the choice of
parameters is meaningful for describing the physical system. Based on the obtained phase
diagram and on the above considerations, a single value of the lattice
spacing, set by $\beta = 2.05$, was chosen, and bare fermion masses were
considered in the range $-1.523 \le am \le -1.475$. For the
quantitative measurements that follow, lattice sizes of $N_T\times
N^3$ between $16\times8^3$ and $48\times{24}^3$ were
considered. Lattice volumes and other parameters are shown in
Tab.~\ref{tab:lattices}.  

\subsection{Centre symmetry}
\begin{figure}
	\center
	\subfloat[C6]{
		\includegraphics[width=\columnwidth]{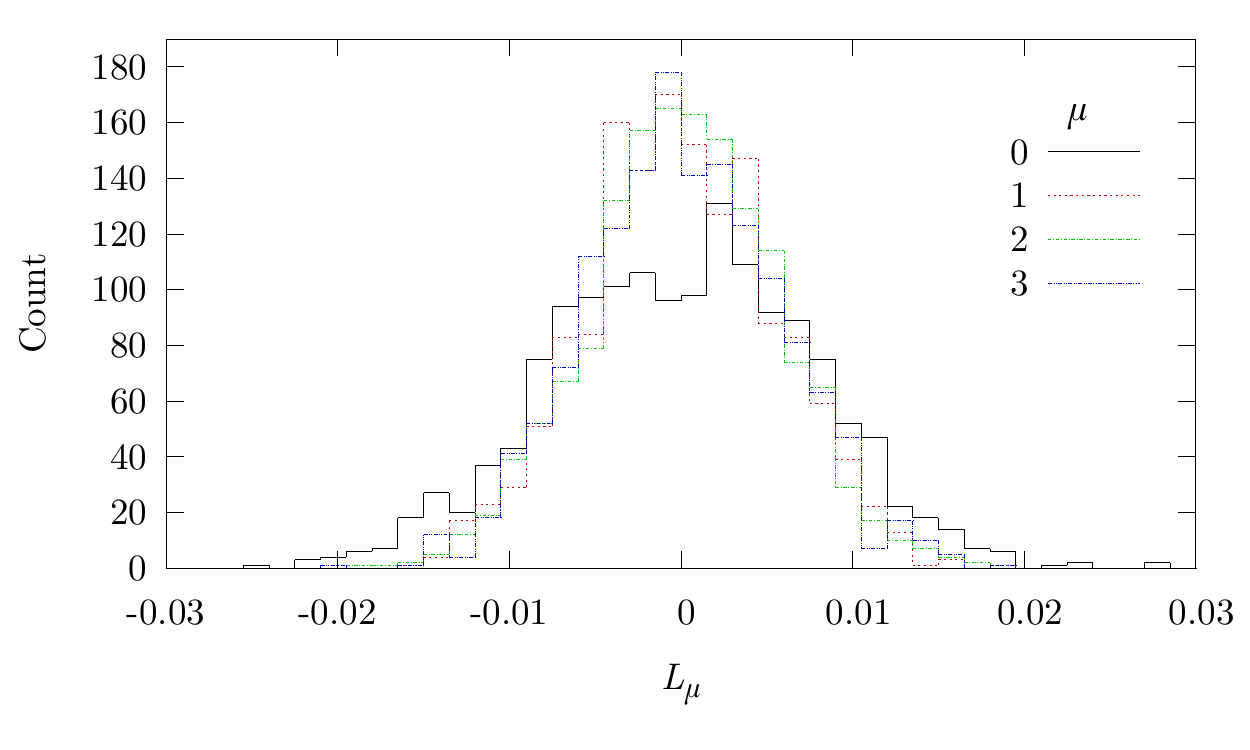}
	}
	
	\subfloat[D1]{
		\includegraphics[width=\columnwidth]{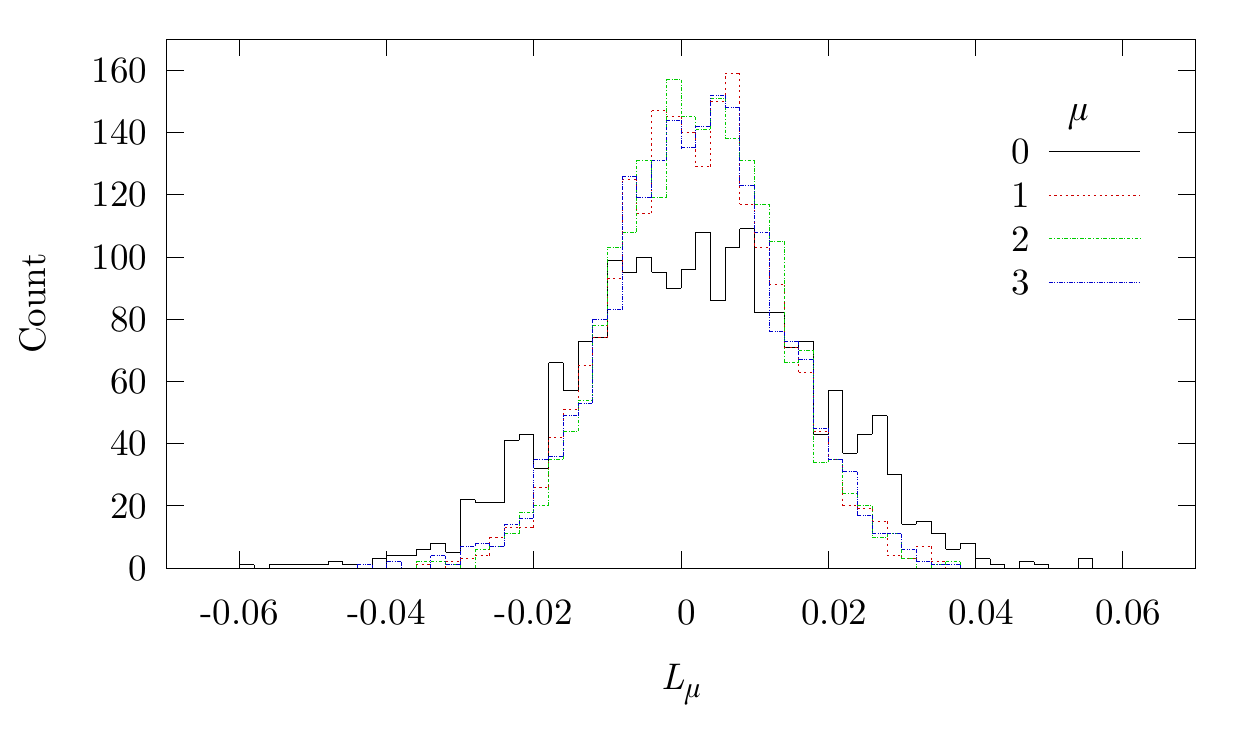}
	}
	
	\subfloat[D2]{
		\includegraphics[width=\columnwidth]{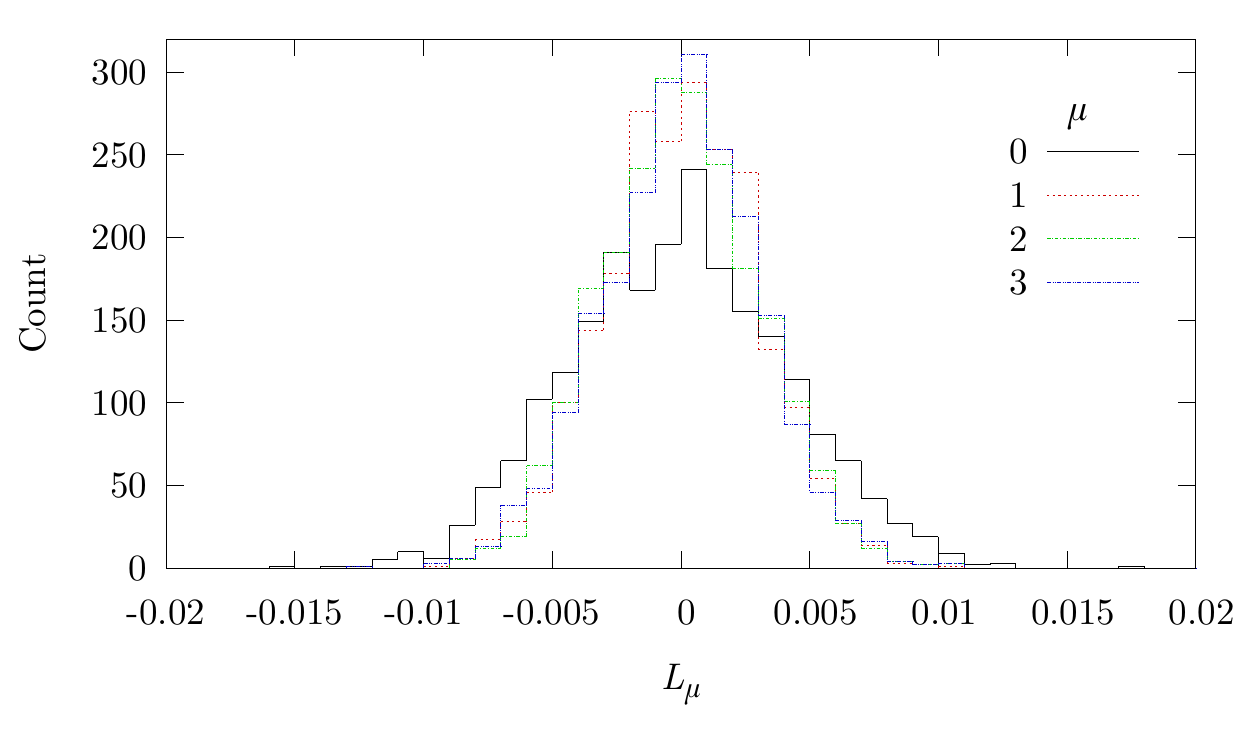}
	}
	
	\caption{(Colour online) The histogram of the average Polyakov loop for all
          configurations belonging to the set shown in the label of each
          subfigure, for all space-time directions. The single peaks
          indicate an unbroken centre symmetry.} 
	\label{fig:centresymmetry}
\end{figure}

 At zero temperature and infinite spatial volume, SU($N$)
gauge theories with adjoint fermions preserve the $(\mathbb{Z}(N))^4$
symmetry related to centre transformations in the four Euclidean
directions. When shrinking the volume or increasing the temperature (the
two mechanisms being connected in an Euclidean setup\footnote{As in
  our simulations, we have assumed periodic boundary conditions for
  the gauge fields in all directions, periodic boundary conditions
  for fermionic fields in spatial directions and antiperiodic boundary conditions for fermionic fields
  in the temporal direction, the latter being related to the finite
  temperature setup.}), the system can pass through various phases (or more precisely
regimes, if the number of degrees of freedom is finite) with different
centre symmetry patterns~\cite{Cossu:2009sq}.

The order parameter for the centre symmetry factor associated to the
direction $\hat{\mu}$ is the vacuum expectation value $\langle
L_{\mu} \rangle$ of the traced Polyakov loop in that direction
\begin{eqnarray}
L_{\mu} = \sum_{i_{\perp}} \mathrm{Tr} \left(\prod_{i=0}^{N_{\mu}} U_{\mu}(i_{\perp},i)\right)
\ ,
\end{eqnarray}
where $i$ is the coordinate ${\mu}$-th coordinate, $i_{\perp}$
the set of coordinates in the perpendicular directions to $\hat{\mu}$
and $N_{\mu}$ the number of lattice points in the $\hat{\mu}$
direction.  

The Polyakov loop can be used to detect finite volume artefacts. On a
sufficiently large lattice, the distribution of the vacuum expectation values of all four
Polyakov loops are symmetric with a peak at zero. A change of regime will occur
when, as reducing the lattice size, $N$ peaks will start to appear in one
of the Polyakov loop distributions. As the lattice size is further reduced,
the Polyakov loops show a more complicated pattern characterised by
the distribution of one or more of them having $N$ peaks. Finally, in
the zero-volume limit, all the four Polyakov loop distributions are
again peaked
at zero~\cite{Cossu:2009sq}. Following the order of the discussion
above, the regime connected with the infinite
volume limit the first of the two regimes characterised by a maximum of
the Polyakov loop histogram at zero. In order to disentangle this from
the opposite zero-volume limit, the Polyakov loop average needs to be
investigated as a function of the lattice size.

The regime of our ensembles with respect to the $(\mathbb{Z}(2))^4$
symmetry  was tested by investigation of averages of Polyakov
loops. When finite size effects are absent, we expect the 
average of traced Polyakov loops in all directions to be consistent with zero
within errors, with the histogram of values having a single peak at
zero and being symmetric around that value. Polyakov loop averages with
a  doubly-peaked distribution at two non-zero values symmetric
around the origin signal finite size
effects~\cite{DelDebbio:2010hx}. Representative plots of 
$\langle L_{\mu} \rangle$ showing a distribution peaked in zero are given in
Fig.~\ref{fig:centresymmetry}.  All the results reported  here have
been obtained for choices of parameters for which the system is in a
regime with all the distributions of Polyakov loops peaked at zero. In
order to check that this is the regime of the thermodynamic limit, we
have observed that reducing the lattice size below a certain value
gives rise to a two peak structure developing in the Polyakov
loop distributions. This study indicates that our simulations are free from
centre-related finite size artefacts.

\subsection{Topological charge}
\begin{figure}
	\subfloat[C1] {
		\includegraphics[width=\columnwidth]{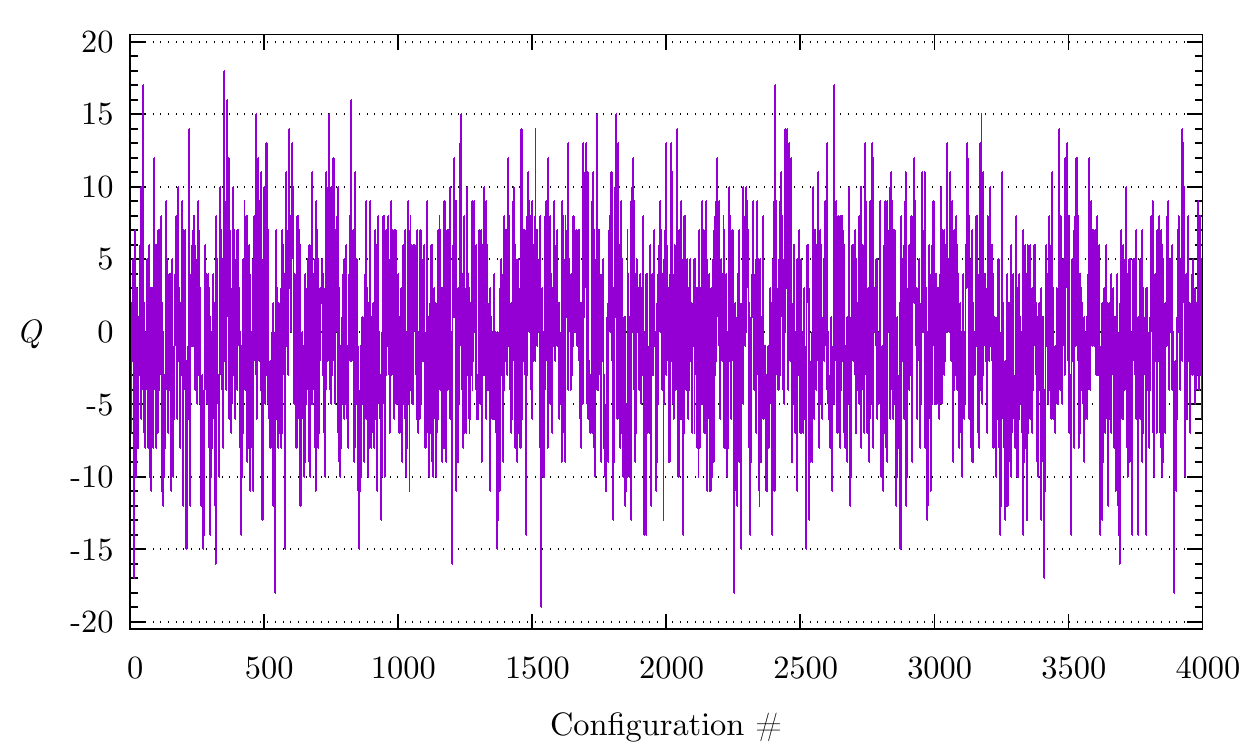}
	}
	
	\subfloat[C4] {
		\includegraphics[width=\columnwidth]{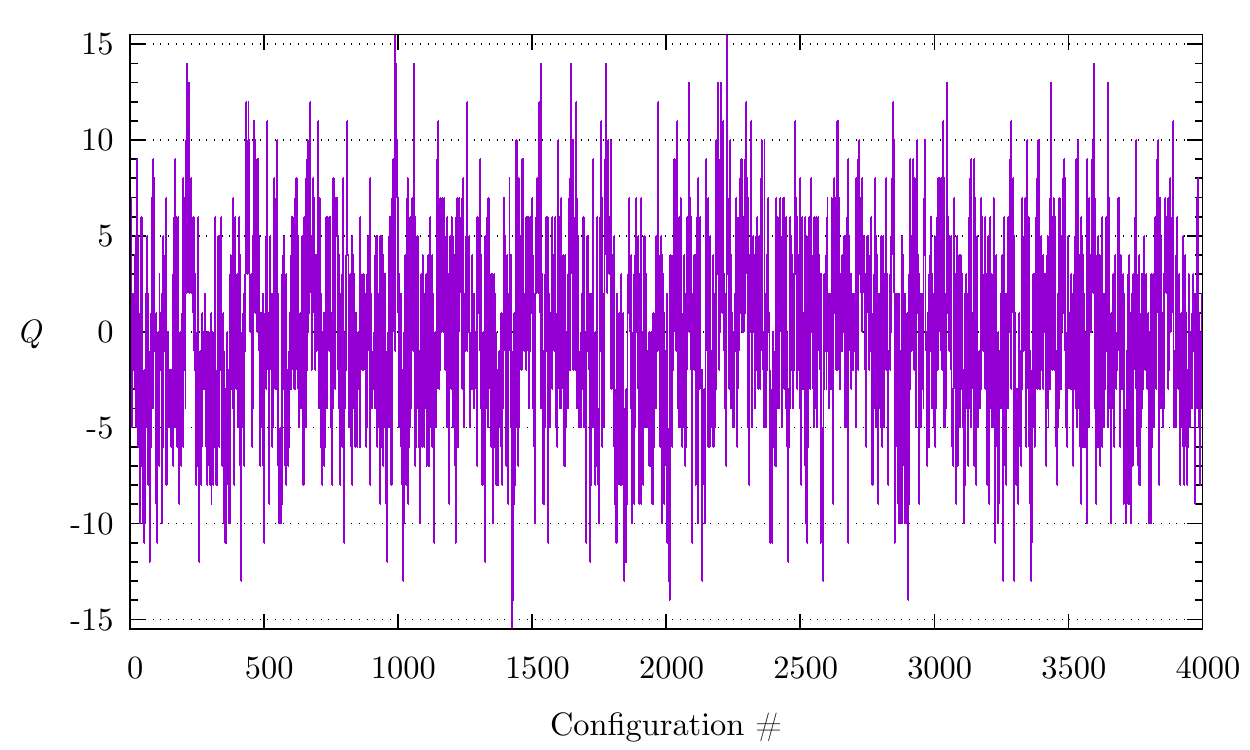}
	}
	
	\subfloat[D2] {
		\includegraphics[width=\columnwidth]{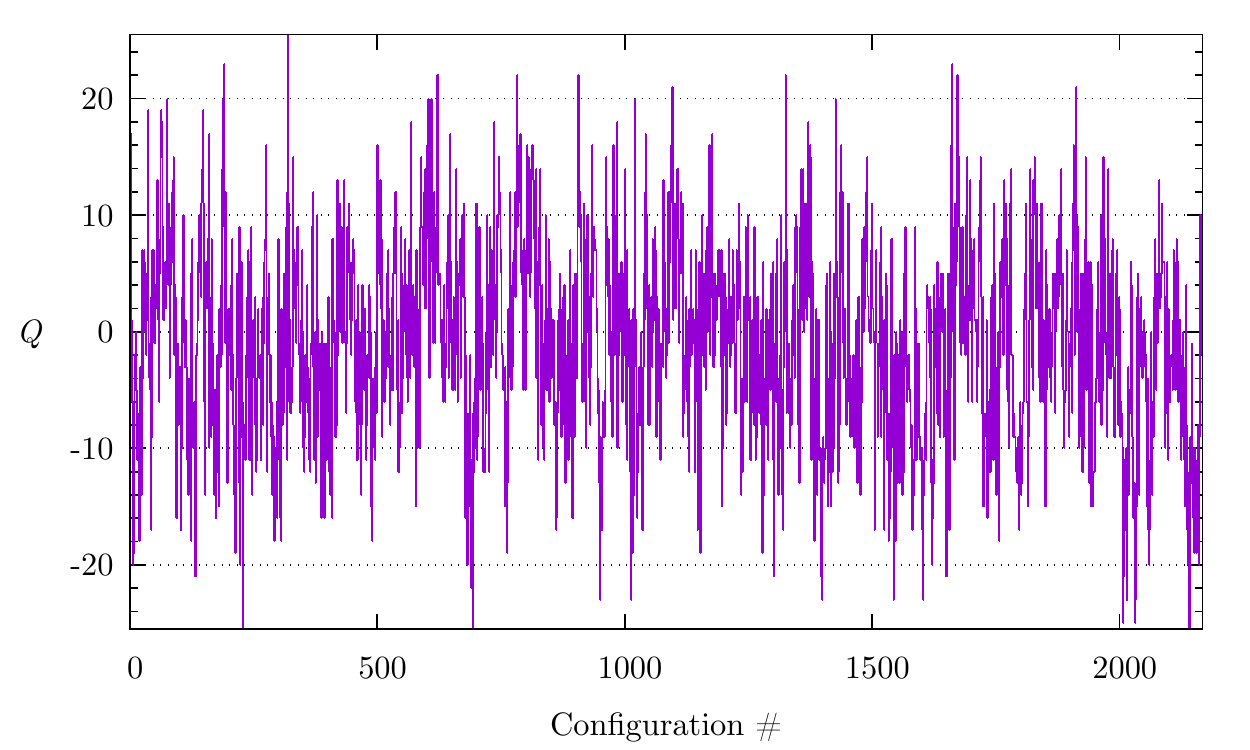}
	}
	
	\caption{(Colour online) The history of the topological charge $Q$ for the indicated ensembles; other ensembles had similar properties.\label{fig:Qhistory}}
\end{figure}
A potential problem of lattice simulations of gauge theories is the emergence of strong
correlations among topological sectors at couplings that are crucial
for taking the continuum limit. In order to understand whether this
also happens in our case, we have monitored the topological charge
history of our runs. 

The topological charge of a lattice configuration can be defined as
\begin{eqnarray}
\label{eq:topcharge}
Q = \frac{1}{32 \pi^2} \sum_i \epsilon_{\mu \nu \rho \sigma} \mathrm{Tr}
\left( U_{\mu \nu}(i) U_{\rho \sigma}(i) \right) \ ,
\end{eqnarray}
where the sum extends over the whole, $\epsilon_{\mu \nu
  \rho \sigma}$ is the fully antisymmetric tensor and $U_{\mu \nu}(i)$
is the plaquette starting from point $i$ in direction $\hat{\mu}$ and coming
back to $i$ from $i +\hat{\nu}$ . In the continuum limit, $Q$ is an
integer labelling the topological sector to which the configuration
belongs.  

Monte Carlo determinations of $Q$ are hindered by ultraviolet
fluctuations, which hide the underlying topological structure. These
fluctuations can be removed using smoothing techniques such as
cooling~\cite{Teper:1985rb} or the more recently introduced Wilson
flow~\cite{Luscher:2010iy}, which are expected to provide similar
benefits~\cite{Bonati:2014tqa}. 

In an ergodic simulation, the system should efficiently explore
topological sectors. However, strong correlations are shown to appear
when the continuum limit is approached.
These correlations determine an increase of the required number of
configurations that are needed to obtain statistically significant
vacuum expectation values of physical observables. A recent
description of the problem for QCD (and a proposed
solution) can be found in~\cite{Luscher:2011kk}. Similar
autocorrelations have been observed in investigations of novel strong
dynamics beyond the standard model (see e.g.~\cite{Fodor:2014cpa}).  

We have measured the value of the topological charge for our configurations using
equation~(\ref{eq:topcharge}). The ultraviolet fluctuations were filtered
out using the cooling method described
in~\cite{Smith:1998wt,Teper:1985rb,Bennett:2012ch}. Representative 
sample histories of $Q$ are shown in Fig.~\ref{fig:Qhistory}. In general, the
results were found to show good tunnelling behaviour, confirming that
the Monte Carlo was not trapped in a single topological sector and
supporting the robustness of our error estimates.

Another check of finite size artefacts is provided by a study of the instanton size
distribution~\cite{Bennett:2012ch}. For our choice of parameters, the
instanton size distributions are those expected in the large volume
limit. This investigation provides another indication that our simulations
are free from the most obvious finite size artefacts.

\begin{figure}
	\center
	\includegraphics[width=\columnwidth]{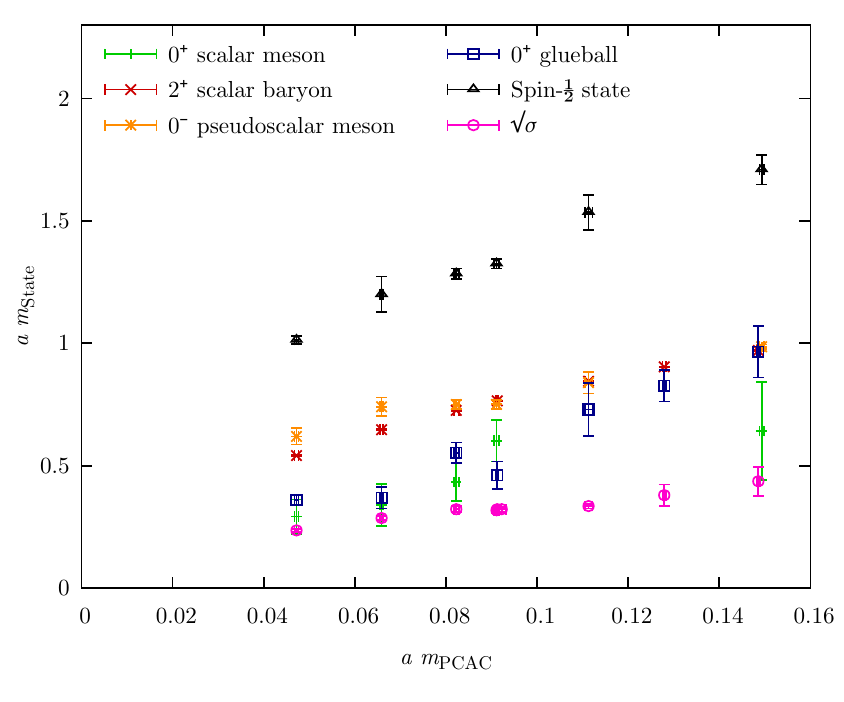}
	
	\vspace{-18pt}
	\caption{(Colour online) Selected spectrum of the theory, showing meson, baryon, glueball, and \spinhalf states, and $\sigma^{1/2}$. \label{fig:spectrum-unscaled}}
\end{figure}
\begin{figure*}
	\subfloat[Mesons and the $0^{+}$ glueball.]{
		\includegraphics[width=\columnwidth]{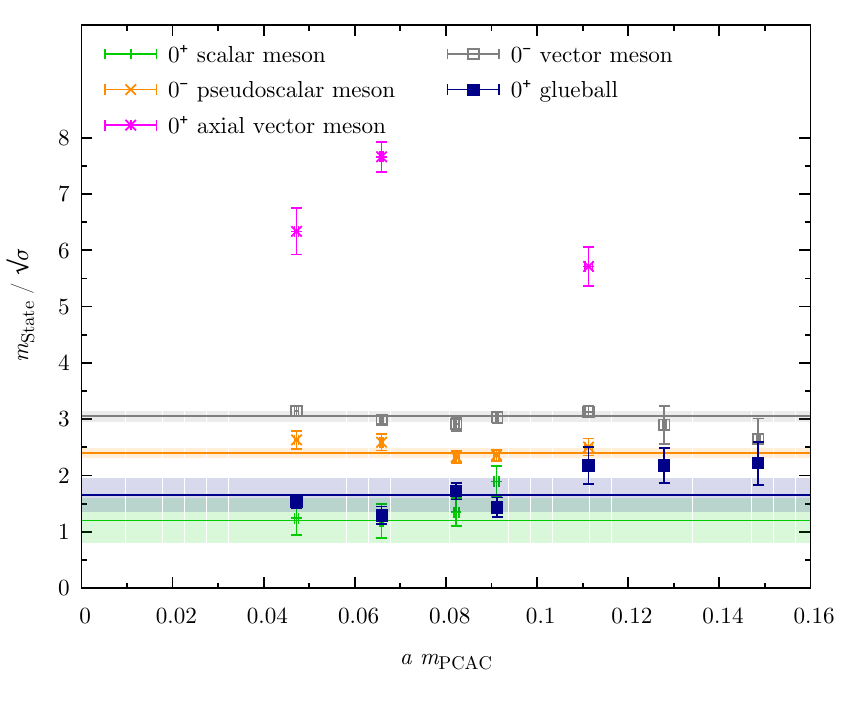}
	}
	\hfill
	\subfloat[Baryons and the \spinhalf state.]{
		\includegraphics[width=\columnwidth]{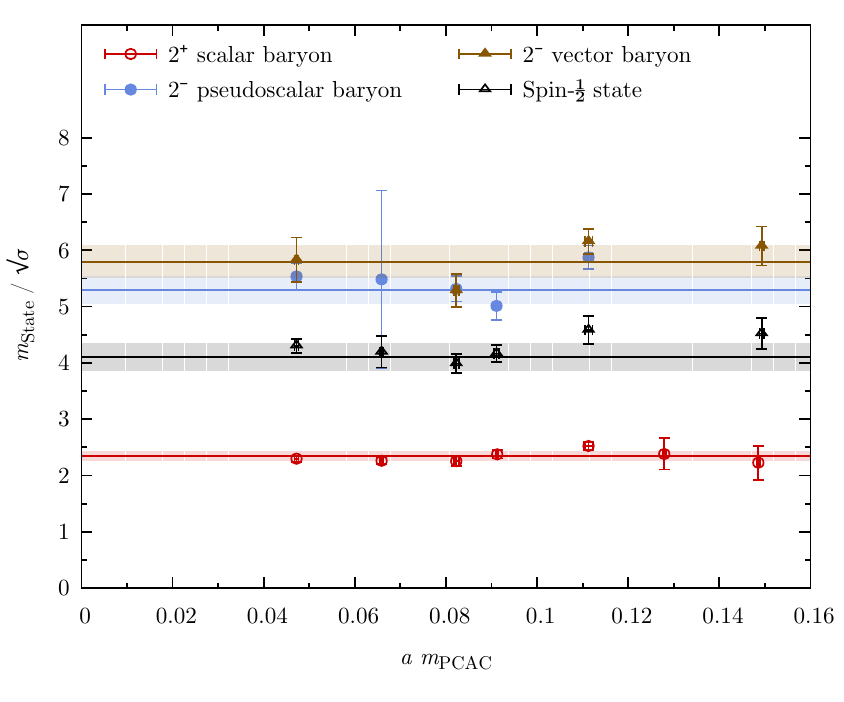}
	}
	
	\caption{(Colour online) Spectrum of the theory, showing meson, glueball, and \spinhalf states, all normalized by $\sigma^{1/2}$. \label{fig:spectrum}}
\end{figure*}

\subsection{Spectral observables}
For the spectroscopic study, we considered masses of mesonic and
baryonic two-fermion states, the $0^{+}$ glueball (the $2^{+}$ was
also considered, but found to be too noisy to provide useful information), and a \spinhalf state,
as well as the fundamental string tension. Baryonic observables were
calculated using two codes: one (HiRep) working in the Dirac basis and using
the Z2SEMWall smearing method~\cite{Boyle:2008rh}, and one developed
for lattice studies of Super Yang--Mills
theories~\cite{Bergner:2013nwa}. Mesonic observables were calculated 
using the latter  code. Gluonic observables were calculated using the
methods described in~\cite{Lucini:2010nv,DelDebbio:2010hx}. The
\spinhalf state is constructed in the continuum from the operator 
\begin{equation}
	O_{\textnormal{spin-}\frac{1}{2}} = \sum_{\mu,\nu} \sigma_{\mu\nu} \tra [F^{\mu\nu} \xi]\;,
\end{equation}
where $\sigma_{\mu\nu} = \frac{1}{2}[\gamma_\mu, \gamma_\nu]$. This state,
which can be seen as a bound state of a fermion and a gluon, was
computed using the tools described in~\cite{Bergner:2013nwa}.

\begin{figure}
	\includegraphics[width=\columnwidth]{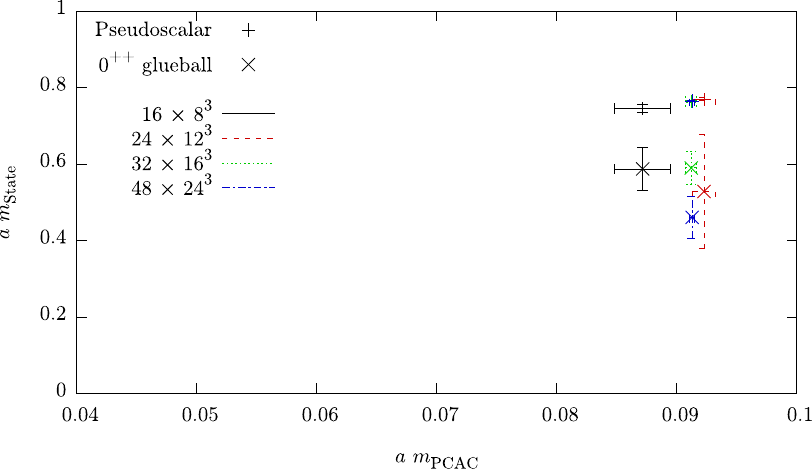}
	\caption{(Colour online) The pseudoscalar baryon and $0^{+}$
          glueball masses, at $a m=-1.51$, for the four lattice sizes considered. \label{fig:finitesize}}
\end{figure}

The spectrum of the theory in lattice units (reported in
Tabs.~\ref{tab:glue},~\ref{tab:mesons}~and~\ref{tab:singlet}) is shown in
Fig.~\ref{fig:spectrum-unscaled}, with Fig.~\ref{fig:spectrum}
displaying ratios of masses over the string tension vs. $\mpcac$.  
In these figures, where multiple lattice volumes were used at a single
fermion mass, the result of the largest was taken. Some states have
proven to be numerically hard to measure, giving large error bars; for
the sake of clarity, those states have been omitted from the figures.

While all masses decrease monotonically
as $\mpcac \rightarrow 0$, the ratio of spectral
quantities to the string tension (determined through correlation
functions of Polyakov loops~\cite{DelDebbio:2010hx}) remains roughly constant for
most quantities in the range studied\footnote{We remark
that after performing a simple extrapolation
to zero $\mpcac$ of our data, most of the quantities discussed in this section are compatible with
zero in the chiral limit. Establishing if all, some, or none of these
spectral quantities are zero in the zero-mass limit is of crucial
importance for a definite answer to the questions we are
addressing. However, this would require more extensive studies on
larger lattices, which are outside the scope of this work.}, with some (particularly the
scalar glueball) showing some deviation (albeit within two standard deviations) at
large fermion mass. Similar behaviour is observed for the $0^+$
meson mass, which is compatible with the mass of the scalar glueball.
A straightforward interpretation of these results is that the glueball
set of operators and the meson operator with $0^+$ quantum numbers
project onto the same ground state. This provides support for mixing
between the scalar meson and the scalar glueball. 

We also note that
this scalar state is the lightest state in the spectrum. After having been
observed in~\cite{DelDebbio:2009fd} in SU(2) gauge theory with
adjoint fermions, the presence of a light scalar has proven to
be a feature that keeps recurring in gauge theories near the conformal
window (see~\cite{Aoki:2013zsa,Aoki:2014oha,Fodor:2014pqa} for recent
lattice investigations and~\cite{Erdmenger:2014fxa} for an approach based on 
gauge-string duality). The light scalar,
which might be a signature in this class of models, is suggestive of a
light Higgs appearing in this framework. This result indicates that
strongly interacting dynamics beyond the standard model is not
incompatible with the experimental constraint that the Higgs must be
light in any extension of the standard model. 

To probe the extent of possible finite-size effects, studies of the
pseudoscalar and of the scalar mass (extracted in the gluonic channel)
were made at $a m=-1.51$ at each of the four lattices considered. As is
shown in Fig.~\ref{fig:finitesize}, the results on the three largest lattices were
all found to agree, while the results on the smallest lattice  (of
volume $V = 16 \times 8^3$) are inconsistent with the others. From this,
the requirement that $L\sqrt{\sigma}$ be greater than or equal to its
value for the smallest consistent case (giving $L\sqrt{\sigma} \geq
3.8$) identifies the configuration sets that are free from finite size
artefacts in the spectrum. The results shown above have been obtained
on sets that respect this constraint. 

\begin{figure}
	\includegraphics[width=\columnwidth]{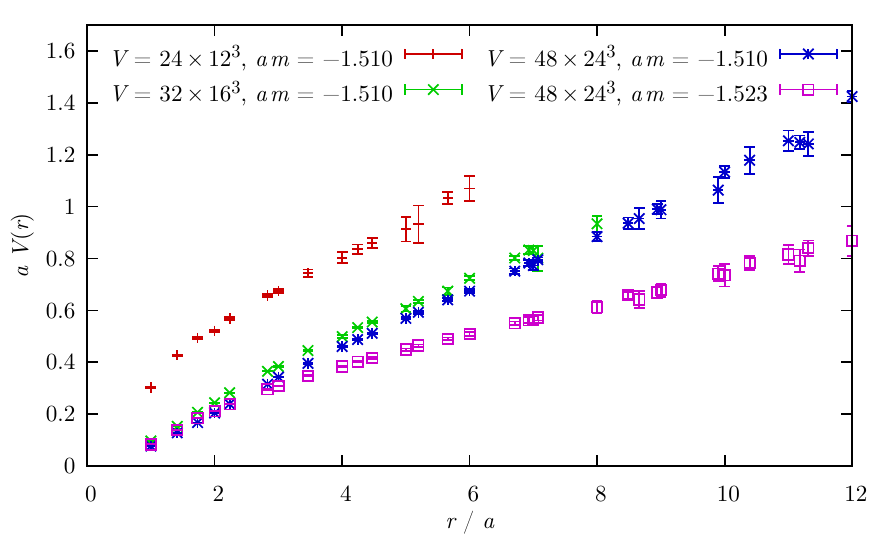}
	\vspace{-8pt}
	\caption{(Colour online) The static fermion potential of four selected lattices. \label{fig:wilson}}
\end{figure}

As a cross-check of the determination obtained from Polyakov loop
correlators, the string tension has been computed also using 
Wilson loops. The technicalities of this calculation are described
in~\cite{DelDebbio:2010hx}. Some examples of the static inter-fermion
potential from Wilson loops are shown in Fig.~\ref{fig:wilson}.  The
string tension has beem extracted using the Cornell ansatz for the
potential. The results found agree with the 
determination obtained using Polyakov loop correlators, but are in
general less accurate. For this reason, in this section we have used
the determination obtained from Polyakov loops.

\subsection{Chiral condensate anomalous dimension}
The flatness of ratios of spectral quantities in this theory as the
fermion mass is send to zero are suggestive of the model being in a
conformal or near-conformal
phase~\cite{DelDebbio:2009fd,DelDebbio:2010hx,DelDebbio:2010hu}.
This indication is reinforced by the fact that the scalar is the
lightest state in the theory. We stress that at the current stage of
our investigations numerical evidence should be taken only as a
hint. With this in mind, a conventional chirally broken phase
seems to be excluded\footnote{As we will remark more in detail in the
  following section, the data are also compatible with a {\em walking}
  behaviour, i.e.\ a phase that at intermediate energies look
  infrared conformal, but at lower energies is in fact chirally breaking.}. For this
reason, for the purpose of understanding quantitatively the character
of the theory, we will use a conformal ansatz, disregarding chiral
perturbation theory as a possible explanation of our model. This
choice is based on the observation that in chiral perturbation theory
the lightest degree of freedom is to be found in the $2^+$ baryon
channel. Hence, even if the system exhibited chiral symmetry breaking,
our data would not be in the 
asymptotic regime,  with the consequence that chiral perturbation
theory is not applicable in this range of masses. Note that similar
considerations also hold if one include the light scalar in the
effective lagrangian as done in~\cite{Matsuzaki:2013eva}, the $2^+$ baryon still
being the only degree of freedom surving in the deep infrared.

Near-conformality can be exploited to determine the value of the
anomalous dimension of the chiral condensate $\gamma_*$.
We may use various techniques to extract $\gamma_*$. We have used two
methods for this work: finite-size scaling predictions and scaling of
the spectrum of the Wilson Dirac operator. For the first case, for a
conformal theory, as a function of the $\mpcac$ mass, a spectral quantity $m_{\mathrm{X}}$ of the system
follows the scaling
relation~\cite{DeGrand:2009mt,DeGrand:2009hu,Lucini:2009an,DelDebbio:2010hx,DelDebbio:2010hu,DelDebbio:2010ze,DelDebbio:2010jy}  
\begin{equation}
	L a m_{\mathrm{X}} = f \left(L\left(a\mpcac\right)^{\frac{1}{1+\gamma_*}} \right)\;,
\end{equation}
for some function $f$, where $L\rightarrow \infty$ is the finite
spatial extent of the lattice, and the combination
$L\mpcac^{1/(1+\gamma_*)}$ is kept constant. If the system is in the
scaling region, then this relation may be used to estimate $\gamma_*$
in the following way. 

\begin{figure}
\begin{center}
	\includegraphics[width=\columnwidth]{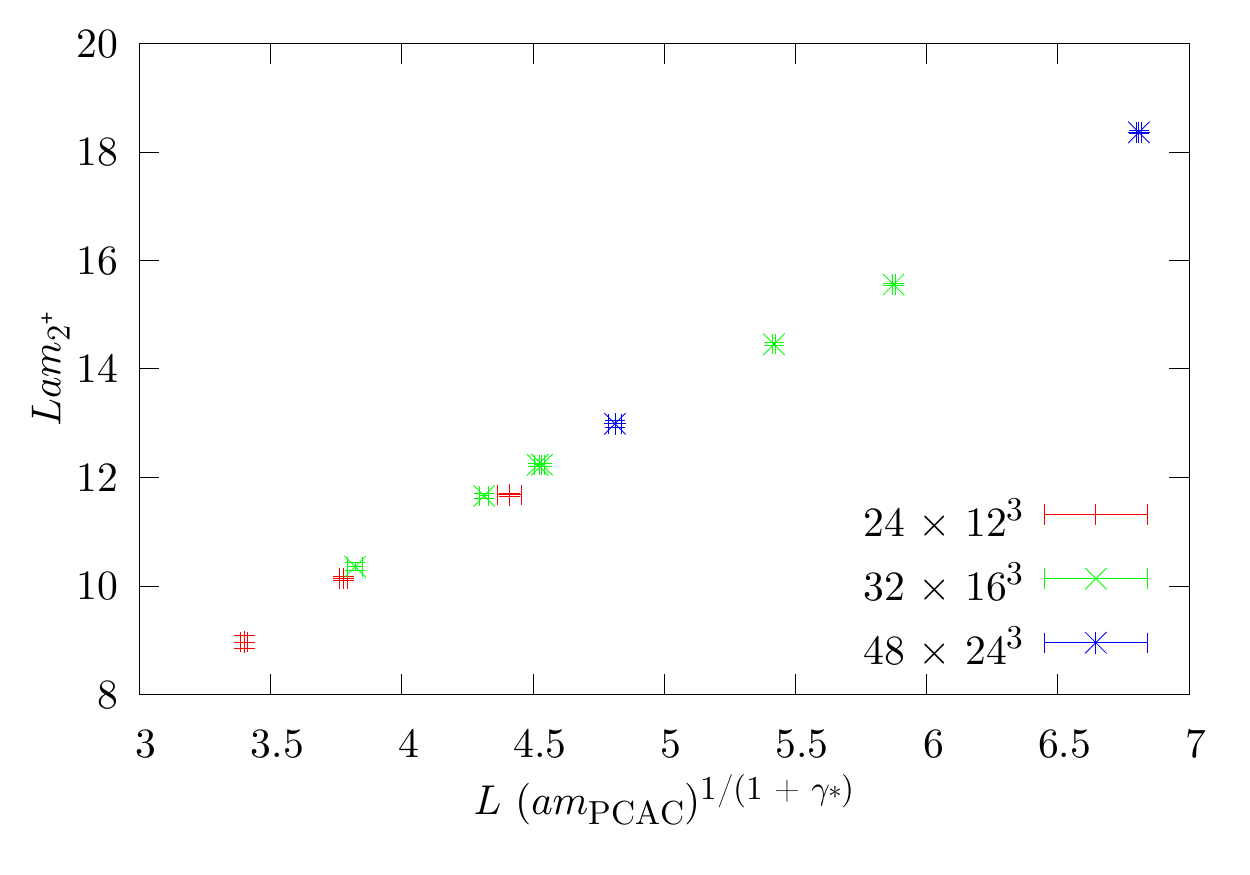} 
	\includegraphics[width=\columnwidth]{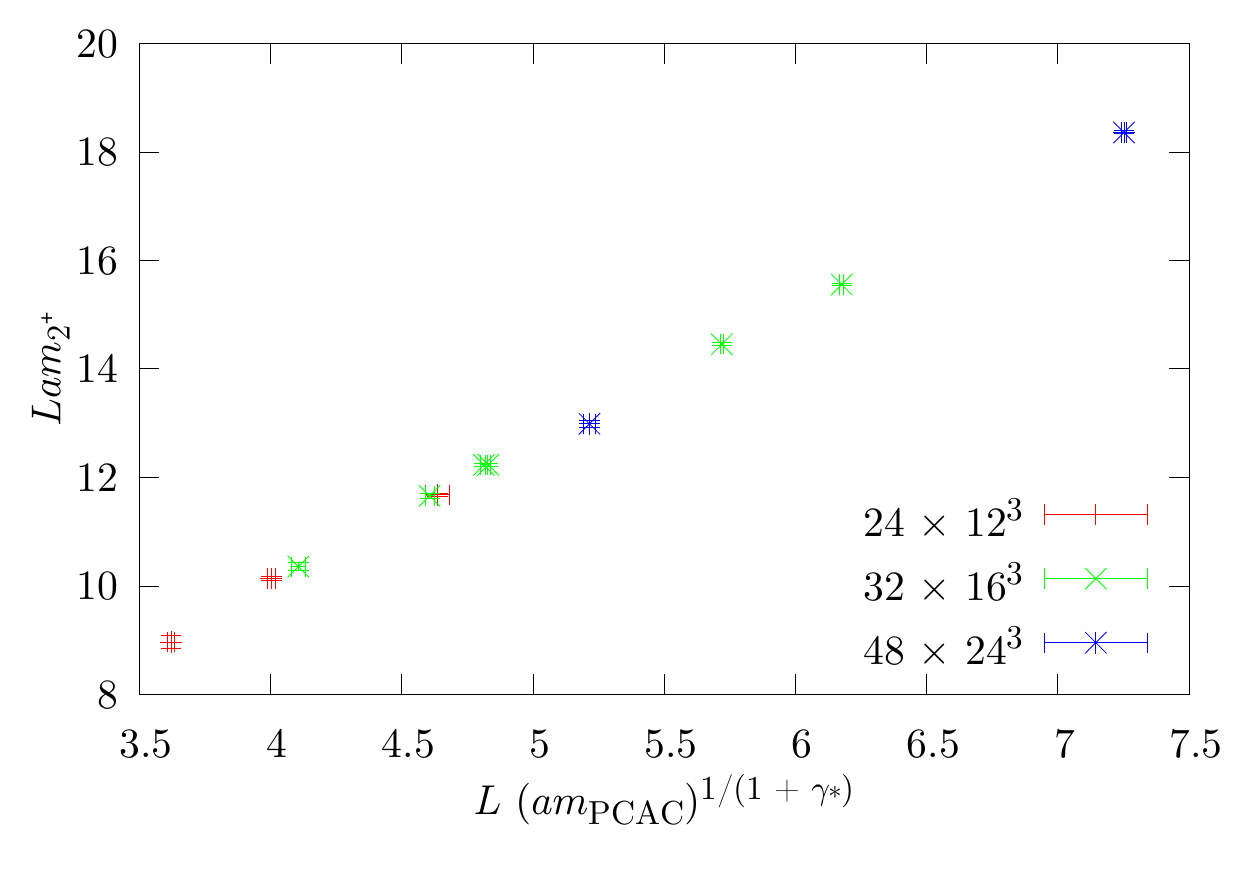} 
	\includegraphics[width=\columnwidth]{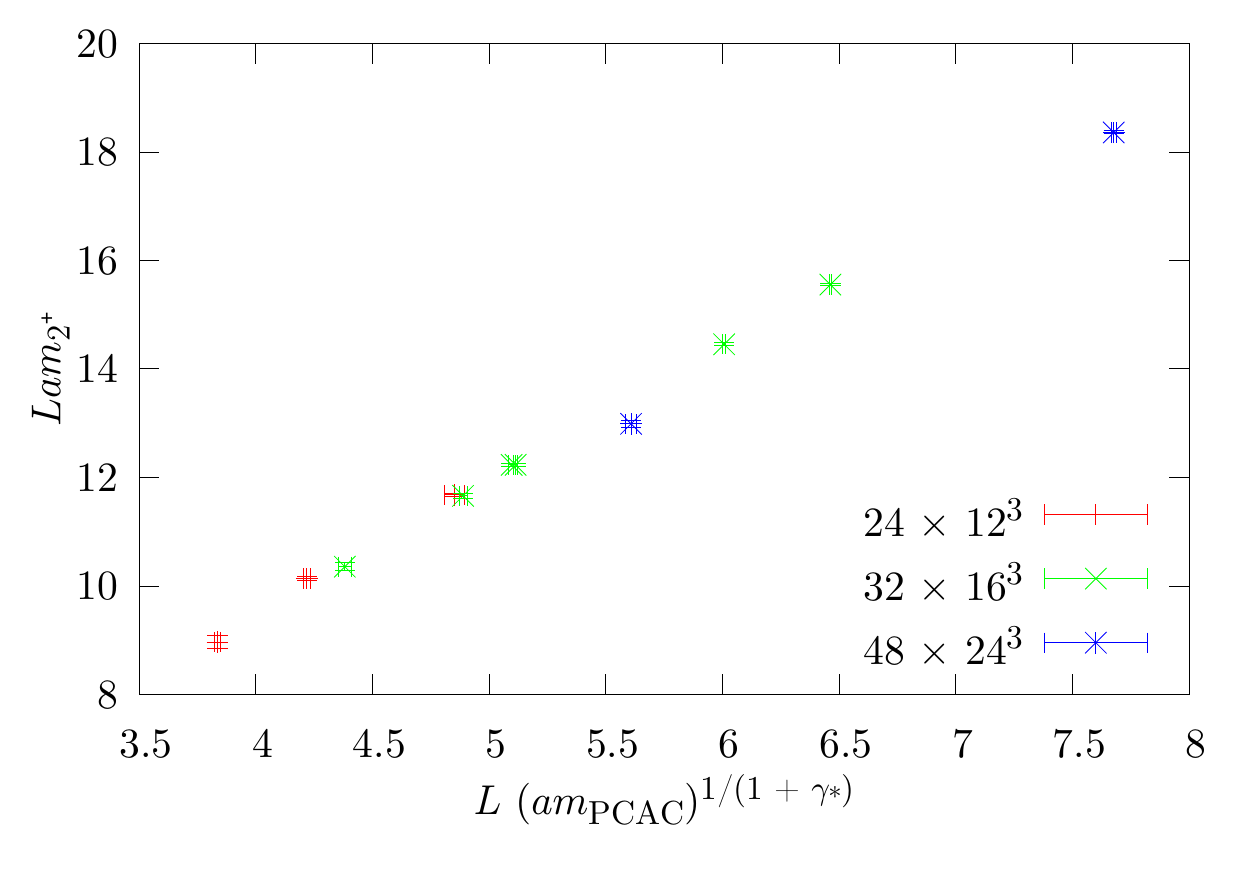}
  \end{center}
\vspace{-0.60cm}
\caption{(Colour online) Plots of $L a m_{{2^+}}$ as a function of $L a \mpcac^{{1}/({1+\gamma_{\ast}})} $ for the three lattice volumes $24\times12^3$, $32\times16^3$ and $48\times24^3$ and $\gamma_{\ast}=0.9, \ 1.0, \ 1.1$. The results appear to identify a universal curve for $\gamma_{\ast}= 0.9 - 1.0$.}
\label{fig:inspection}
\end{figure}

Firstly, we plot $Lam_{\mathrm{X}}$ against $L\left(a\mpcac\right)^{1/(1+\gamma_*)}$
for all available lattice volumes on one plot for each of various
values of $\gamma_*$. We then take the set of plots and find the
region of $\gamma_*$ that allows the sets from different lattices to
lie on a single universal curve. In Fig.~\ref{fig:inspection} we see
the $2^+$ scalar baryon analysed in this manner, at three values of
$\gamma_*$; we see the best fit is observed in $0.9\le\gamma_0\le1.0$,
and so we expect the anomalous dimension to lie in this
region\footnote{While a similar analysis on other states give
  compatible results, the $2^+$ baryon, being the most accurately
  determined state in the spectrum, allows us to perform a better
  determination of $\gamma_*$.}.

\begin{figure}
	\includegraphics[width=\columnwidth]{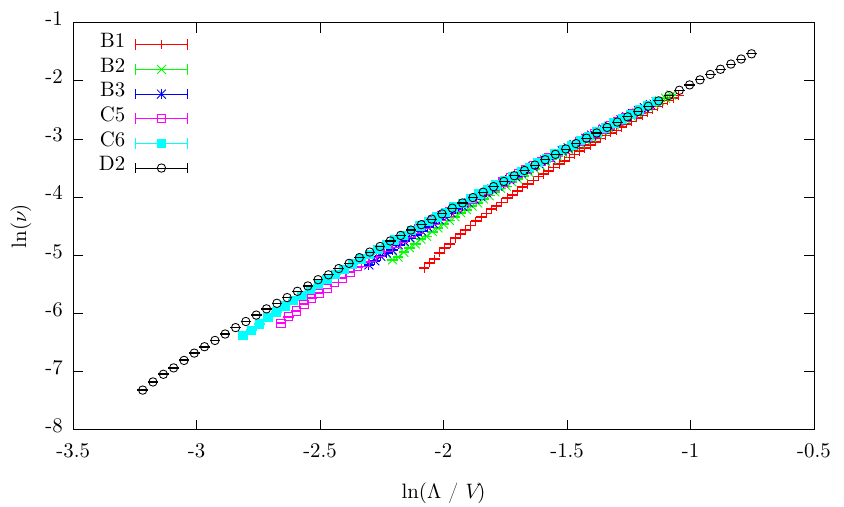}
	
	\vspace{-8pt}
	\caption{(Colour online) The behaviour of the Dirac mode number for a subset of the lattices considered. \label{fig:modenumber}}
\end{figure}
\begin{figure}
	\includegraphics[width=\columnwidth]{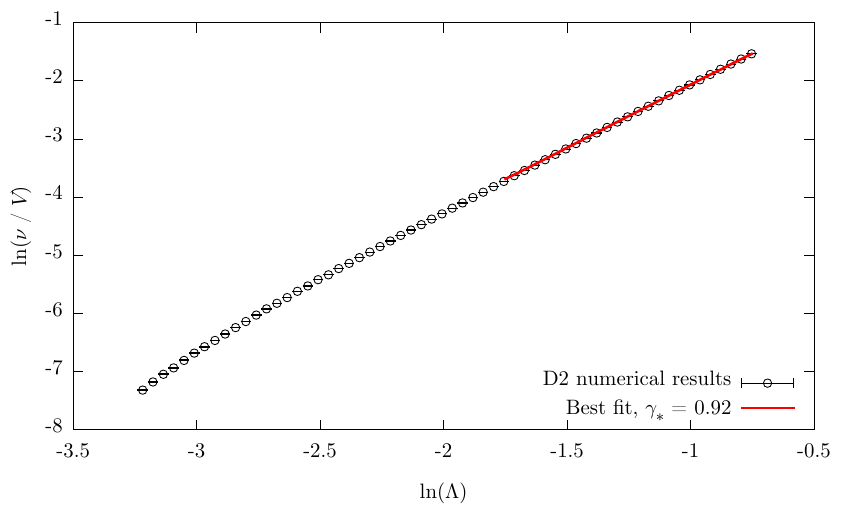}
	
	\vspace{-8pt}
	\caption{(Colour online) The behaviour of the Dirac mode number on the D2 lattice, compared with the results of the numerical fit, with $\gamma_* = 0.92$.\label{fig:modefit}}
\end{figure}
A more precise method of determining $\gamma_*$ is to fit the Dirac
mode number $\nubar(\Omega)$ as a function of the Dirac eigenvalue
$\Omega$ \cite{Patella:2012da} (see also~\cite{Cheng:2013eu}). We expect the mode number to scale as:
\begin{equation}
	a^{-4}\nubar(\Omega) \approx a^{-4} \nubar_0(M) + A[(a\Omega)^2-(aM)^2]^{\frac{2}{1+\gamma_*}}\;,
\label{eq:nubar-scaling}
\end{equation}
where $M$ is some renormalised fermion mass. The raw output of a set of simulations is plotted in Fig.~\ref{fig:modenumber}.

In numerical studies of a mode number distribution that follows this
relation, we have four parameters to fit for: $\nubar_0(M)$, $A$, $(aM)^2$, and
$\gamma_*$. Additionally, in the presence of chiral symmetry, we would
expect this relation to hold for $\Omega\rightarrow0$; however, since
simulations are performed at finite fermion mass, scaling is only seen in an
intermediate range of $\Omega$, which is not known \emph{a
  priori}. This means that in addition to fitting for the four
variables above, we must also carefully locate the scaling window of
$\Omega$. We choose to perform this analysis on the D2 lattice only,
since the longer extent will provide the greatest opportunity to
observe the scaling region. The corresponding data are plotted in
Fig.~\ref{fig:modefit}. 

To do this, we consider each possible window $[\Omega_{\mathrm{LE}},
\Omega_{\mathrm{UE}}]$ in turn, and perform an initial fit using two
algorithms, Levenberg--Marquadt and simulated annealing, with $m=0$ set
to allow a convergent fit. To obtain an estimate of the stability of
the fit, these initial values are then fed back into the same fitting
algorithm a large number of times, with some random ``jitter''
applied, and bootstrap sampling is used to estimate the error on the
average outputs. 

\begin{figure*}[hbt]
	\vspace{-12pt}Lower end of window:
	\center
	\makebox[\columnwidth][c]{\includegraphics{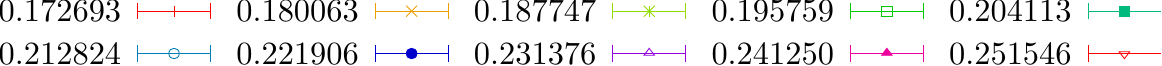}}
	\null
	
        \subfloat{
		\includegraphics[width=\columnwidth]{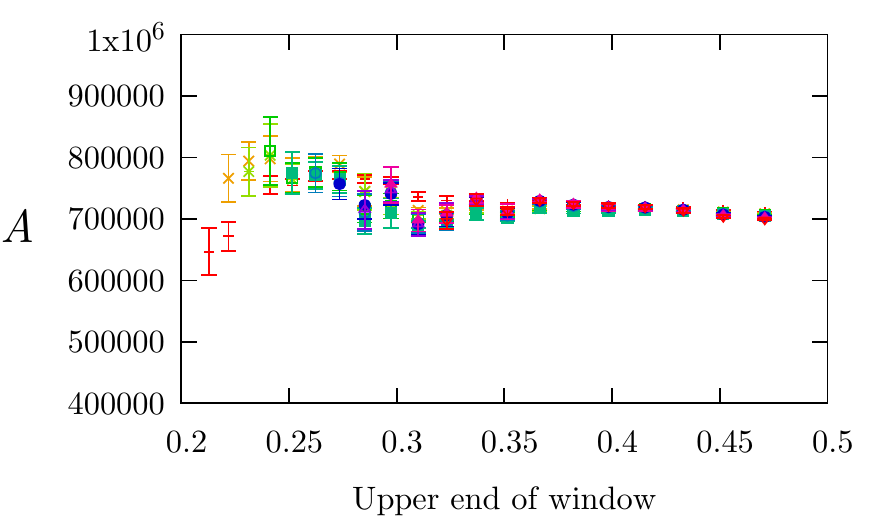}
              }
              \hfill
              \subfloat{
                \includegraphics[width=\columnwidth]{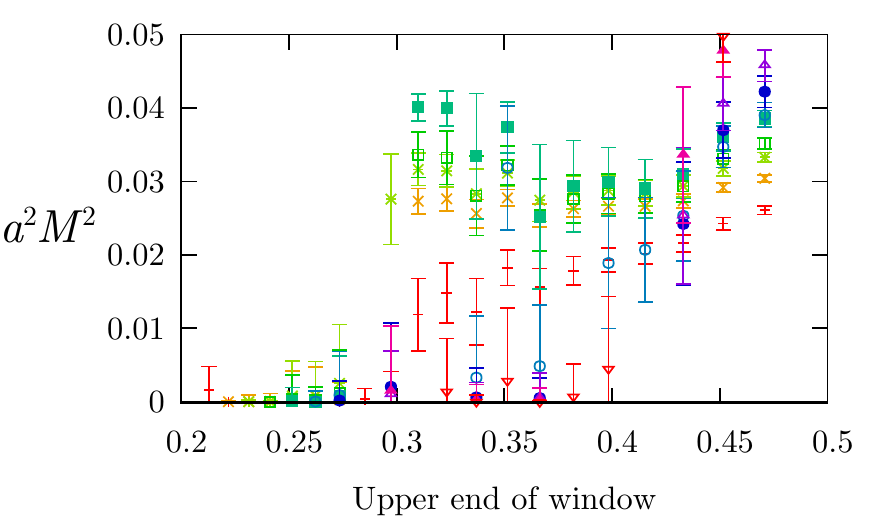}
              }
              ~\\
              \subfloat{
		\includegraphics[width=\columnwidth]{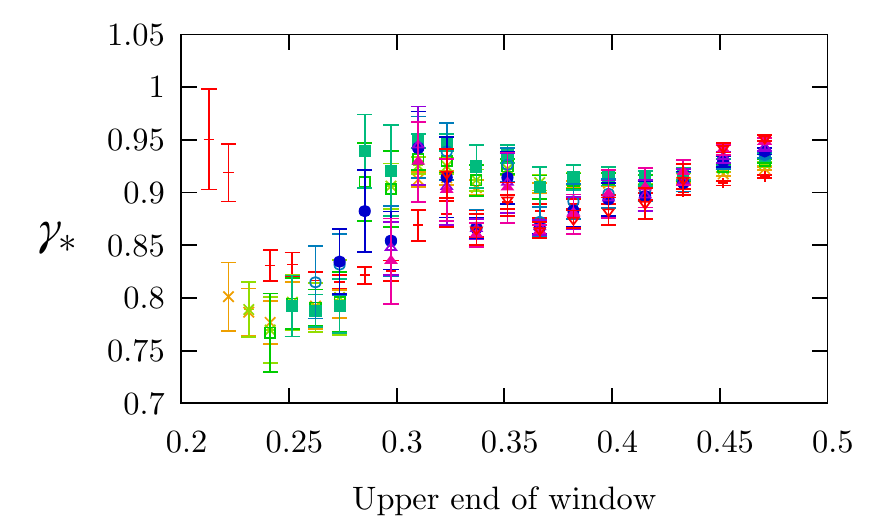}
              }
              \hfill
              \subfloat{
		\includegraphics[width=\columnwidth]{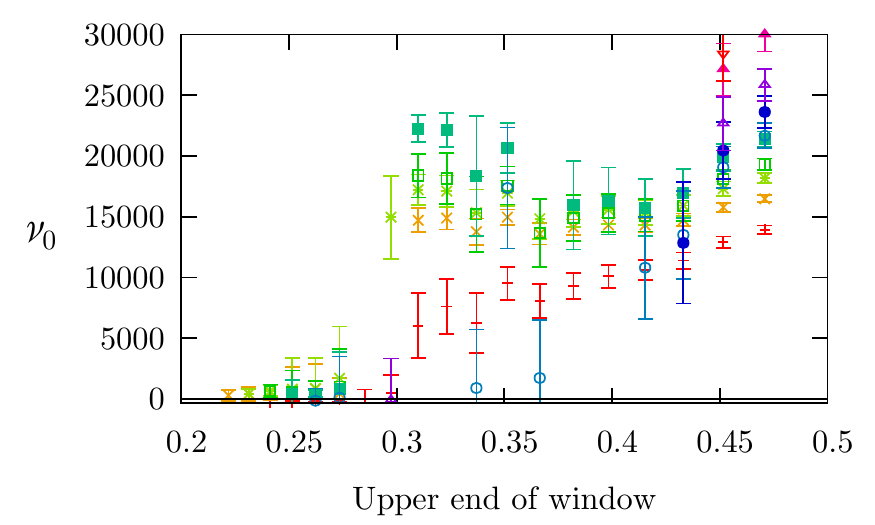}
              }		
	\caption{(Colour online) Plateaux for the fitted observables for the D2 data
          at various lengths and positions of the fitting window. The
          colour represents the position of the lower end of the
          window, and the $x$-axis the upper end. The plateaux at the
          top-right of each plot were taken as the central values of
          our estimates.}
	\label{fig:windows}
\end{figure*}

From this, a set of plots can be drawn for each variable showing the
value and error for all possible windows. Regions of stability in each
variable can be seen as plateaux, and the scaling region is identified
as the region that is most stable on all four plots
simultaneously. This analysis (limited to a subset of
$\Omega_{\mathrm{LE}}$, $\Omega_{\mathrm{UE}}$ for the sake of
readability) for the Levenberg--Marquadt results on the D2 lattice is
shown in Fig.~\ref{fig:windows}, where $\gamma_*$ was consistently
found to lie in the range $0.9\lesssim \gamma_* \lesssim 0.95$, with a
best fit of $\gamma_* = 0.92(1)$, in agreement with the analysis based
on the finite size scaling of $m_{0^-}$. The quality of the best fit
is shown in Fig.~\ref{fig:modefit}. The simulated annealing results were
found to be in good agreement with these data, and the best fit is
consistent with the range found via the spectral scaling
relations. Putting together the Dirac operator eigenvalue and the
spectral scaling determination, a safe estimate for $\gamma_*$ is
$\gamma_* = 0.925(25)$. This is the highest value for a condensate
anomalous dimension found in any lattice simulation. 

\section{Discussion}
\label{sect:discussion}
Taken at face value, the numerical results of the previous section
would imply infrared conformality of the theory with an anomalous
dimension of order one. Since both the infrared conformality of
the theory and the large anomalous dimension are somewhat
unexpected, in this section we review arguments that seem to suggest
a different result, discuss their extent of validity (and the extent
of validity of our analysis) and outline where further simulations
will help in pinning down potential remaining issues.

We have already stressed the large value of the anomalous dimension of
the condensate, which makes this theory unique among those
investigated on the lattice to date. However, we remark that the
anomalous dimension has been obtained at a single lattice spacing,
while the interesting quantity is its value in the continuum limit. In
SU(3) with $\Nf = 12$ fundamental fermion flavours, a strong lattice
spacing dependence of the anomalous dimension has been observed that
can be successfully described by adding subleading corrections to the
dominant scaling behaviour of observables near the infrared fixed
point~\cite{Cheng:2013xha}. Hence, in order to determine the continuum
value of the condensate anomalous dimension, a more extended
investigation involving higher $\beta$ values would be needed. Such an
investigation will enable us to perform a robust extrapolation that
includes scaling violations. In the absence of a more detailed study,
we will make the working assumption that the value of the anomalous
dimension will not change significantly when extrapolated to the
continuum. 

At a superficial sight, cases for a non-conformal behaviour in the theory
discussed in this paper seem to come from arguments based on the two-loop
$\beta$-function~\cite{Dietrich:2005jn} and arguments based on the
Seiberg and Witten solution of ${\cal N} = 2$ super
Yang--Mills~\cite{Seiberg:1994rs}, which is related to our theory by 
an infinite mass deformation for the scalar.  
As noted in the introduction, the behaviour we have found does not
generate any tension with the Seiberg and Witten result for ${\cal N}
= 2$. In fact, the ${\cal N} = 2$ theory undergoes spontaneous
symmetry breaking from gauge group $\su2$ down to $\uone$ at any value
of the Higgs condensate. When a small mass is given to the Higgs
field, this moduli space is lifted and in the two resulting vacua
monopoles condense, giving rise to Abelian confinement via the dual
superconductor mechanism (already advocated before for QCD in
~\cite{tHooft:1977hy}). A deformation retaining supersymmetry allows one to
construct a path leading from ${\cal N} = 2$ to ${\cal N} = 1$
super Yang--Mills, with the latter theory confining. However, the
deformation that takes ${\cal N}=2$ super Yang--Mills to the gauge
theory investigated in this work does not retain supersymmetry, and as
such does not allow to argue about the phase of the theory we are
interested in, starting from the original analytic result. It should
also be stressed that the original ${\cal N} = 2$ super Yang--Mills
theory lives in the Coulomb branch (i.e.\ the infrared limit is
Abelian and conformal) and when deformed with a small mass term for
the Higgs field the resulting theory has Abelian confinement of the
electric charge, which is inherently different from the confinement
observed in non-supersymmetric non-Abelian gauge
theories~\cite{Ambjorn:1999ym}. In other words, in the regime for which the
mass deformation is controlled, the theory one might think of using as
a starting point for arguing about confinement in the $\Nf = 1$
Dirac adjoint theory nowhere confines in the way a consistent
argument would require.

Concerning the $\beta$-function argument, the underlying result is entirely
based on perturbation theory, which is expected to break down when
attempting a description of a strongly coupled infrared fixed
point. This has been observed for $\su2$ with two Dirac flavours in the
adjoint representation, which was expected to be confining (albeit
near the onset of the conformal window) perturbatively, but has been
shown to be infrared conformal by lattice calculations.

\newcommand\Nw{N_{\textnormal{W}}}
Recently, it has been noticed in~\cite{Basar:2013sza} (see
also~\cite{Basar:2014jua}) that there is an
apparent tension between large-$N$ volume reduction (which is expected
to hold in theories with adjoint fermions) and presence of a
Hagedorn-type density of states (which characterise confining
theories). The authors advocate a solution based on an emerging
fermionic symmetry at large $N$. However, on the light of our study,
another possibility is that theories with $\Nw$ Weyl flavours are not
confining when $\Nw > 1$, as there are indications for our case ($\Nw
= 2$) and for the two adjoint Dirac flavour theory ($\Nw = 4$).

However, even if those arguments in favour of confining
behaviour have possible breaches, it is still possible that our lattice
simulations are in an intermediate mass regime and that at lower masses
the model shows confining behaviour\footnote{We note that this is a
  logical possibility for any lattice simulation of this type, and the
 ability to observe in practice the true chiral behaviour in this case
 crucially depends on the separation between the two scales at which the
 plateau in the spectrum is observed. In particular, for phenomenology a large scale
 separation (of a few orders of magnitude) is required. Observing both
 regimes in a theory with such a large scale separation is a prohibitive task for lattice
 simulations, given the current techniques and the current
 computational resources.}. In order to check whether we can see this
change in behaviour, we are performing simulations at larger volumes
and lower masses. If the theory turns out to be confining in the deep
infrared, the implication is that $\su2$ with one adjoint fermion
is walking, i.e.\ near-conformal in an intermediate energy range
before turning confining at some low energy scale. This is the
wanted behaviour for constructing a phenomenologically viable model of
strongly interacting dynamics causing electroweak symmetry breaking,
with the anomalous dimension being in the region of values compatible 
with experimental constraints.  

Finally, our model only contains two Goldstone bosons associated
with the chiral symmetry breaking, while phenomenologically at least
three would be required to account for the standard model electroweak
symmetry breaking. Nevertheless, the system studied here can be
thought of as a sector of a richer theory containing also fermions in
the fundamental representation, known as {\em ultraminimal
  technicolour}, whose phenomenology has been first explored
in~\cite{Ryttov:2008xe}. Our results suggest that a model constructed
along the lines of ultraminimal technicolor could be compatible with
phenomenology. 

\section{Conclusions}
\label{sect:conclusions}
In this paper, we have performed a first numerical exploration of $\su2$ gauge
theory with one Dirac flavour. After investigating the phase structure of the
theory, we have performed an extensive study at a value of the
coupling that we have found to be continuously connected with the
continuum limit of the theory. By studying the scaling of the spectrum
and of the eigenvalues of the Dirac operator as the mass is reduced
towards the chiral limit and the lattice is kept large enough for
finite size artefacts to be under control, we have found indication of
a conformal infrared behaviour with an anomalous dimension $0.9 \lesssim \gamma_* \lesssim
0.95$. If this features are confirmed by more extended simulations aimed at
extrapolations to the continuum limit, the model will provide the
first theory in the conformal window that has an anomalous dimension
compatible with phenomenological constraints. Another possibility is
that the theory is walking, and we are just observing its behaviour at
intermediate energy scales. This case would also be interesting, as
this would result in the first observation of a near-conformal theory
with a large anomalous dimension. Such an observation would be again
of theoretical relevance towards the construction of a
phenomenologically viable explanation of electroweak symmetry breaking
based on a novel strong dynamics beyond the standard model. Although the
system studied here does not possess enough Goldstone bosons to
provide a complete description of electroweak symmetry breaking due to
a novel strong interaction, it might appear as a sector of a theory
that could realise such a mechanism of electroweak symmetry breaking.

\begin{acknowledgments}
We thank A.~Patella for participating to the early stages of this
project, for his suggestions and his critical reading of this
manuscript. We would like to thank A.~Cherman, T.~DeGrand, M.~Lombardo,
E.~Pallante, F.~Sannino and M.~\"Unsal for discussions on various
aspects of this project. BL is indebted with D.~Dorigoni and
T.~Hollowood for
enlightening discussions on ${\cal N}=2$ Super Yang--Mills. Part of this work
was performed when B.L.\ visited the Aspen Center for
Physics (NSF Grant No.\ 1066293), which he thanks for
hospitality. EB thanks the members of the LatKMI collaboration for
helpful discussions and the Kobayashi--Maskawa Institute 
for hospitality during the latest stages of this work. 
Numerical computations were  executed in part on the Blue Gene/P
machine in Swansea University and the ULGQCD cluster in the University
of Liverpool (part of the DiRAC facility supported by STFC), on the
HPC Wales cluster in Cardiff, supported by the ERDF through the WEFO
(part of the Welsh Government), on the BlueGene/Q system at the
Hartree Centre (supported by STFC) and on the BlueGene/Q system at the
University of Edinburgh (part of the DiRAC2 facility supported by
STFC). This project has been supported by STFC (grant ST/G000506/1),
and also by the JSPS Grant-in Aid for Scientific Research (Grant No. \#23-01781 (EB)). 
EB is also partially supported by the JSPS Postdoctoral Fellowships for Foreign Researchers (PE13578). 
\end{acknowledgments}

\appendix
\section{Notations and conventions}
\label{app:a}
In this Appendix, we describe for convenience the notations and the conventions on the
Dirac algebra we have used for deriving the results of Sect.~\ref{sect:model}.
 
In Minkowski space, with the metric tensor $g=$ diag$(+1,-1,-1,-1)$, we choose to use a chiral representation of the Dirac algebra, with
\begin{equation}
	\gamma^{\mu}=\left(\begin{array}{cc}
		0 & \overline{\sigma}_{\mu}\\
		\sigma_{\mu} & 0
	\end{array}\right)\;,
\end{equation}
where in turn
\begin{equation}
	\sigma^{\mu}=\left(\one_2,\vec{\sigma}\right),\quad\overline{\sigma}^{\mu}=\left(\one_2,-\vec{\sigma}\right)\;.
\end{equation}

$\vec{\sigma} =\left(\sigma_{1},\sigma_{2},\sigma_{3}\right)$ is the 3-vector formed from the Pauli matrices
\begin{equation}
  \sigma_{1}=\left(\begin{array}{cc}
		0 & 1\\
		1 & 0
	\end{array}\right),\sigma_{2}=\left(\begin{array}{cc}
		0 & -i\\
		i & 0
	\end{array}\right),\sigma_{3}=\left(\begin{array}{cc}
		1 & 0\\
		0 & -1
	\end{array}\right)\;.
\end{equation} 

and $\one_2$ is the $2\times2$ identity matrix. As usual, we define
\begin{equation}
	\gamma_5=i\gamma_0\gamma_1\gamma_2\gamma_3= \left(\begin{array}{cc}
		\one_2 & 0 \\
		0 & -\one_2
	\end{array}\right)\;.
\end{equation}

Charge conjugation is defined as
\begin{equation}
	\psi_{\mathrm{C}}=C\pb^\tr
\end{equation}
where
\begin{equation}
	C = i\gamma_0\gamma_2=i\left(\begin{array}{cc}
		\sigma_2 & 0\\
		0 & -\sigma_2
	\end{array}\right)\;.
\end{equation}
The properties
\begin{align}
	C^\dagger = C^\tr &= -C \;,\\
	C\gamma_\mu C &= \gamma_\mu^\tr \;,\\
	C\gamma_5 C &= -\gamma_5\;.
\end{align}
easily follow from the definition.

\section{A note on states and parity}
\label{app:b}
Our naming convention for bound states and lattice operators relevant
for this investigation is different from the one used in lattice
simulations of QCD and supersymmetric Yang--Mills theory. In order to
clarify the correspondence, we discuss in this Appendix the
relation between our classification of the spectrum and others found
in the literature, in particular with connection to the mesonic and
baryonic operators we are using.  If one compares the states studied
here to their QCD equivalents, or to other theories beyond the standard
model, it is important to keep in mind the differences in the
conventions we are going to expose.

In this work, fermion bilinears are meson and baryon operators. This is
due to the fact that the gauge group (in our case, $\su{2}$) is
(pseudo-)real. This is similar to the notion of baryons or diquarks in
other investigations of $\su{2}$ gauge theories with  
fermions in the fundamental representation (see
e.g.~\cite{Hands:1998kk} and references therein).
In the notation in terms of two Majorana fermions, all these states
are referred to as mesons. In supersymmetric Yang--Mills theory a
meson is named after its QCD equivalent, which is a flavour
singlet meson. For instance, the scalar meson is called adjoint $f_0$
and the pseudoscalar meson adjoint $\eta^{\prime}$. 

In QCD the triplet $\gamma_5$ meson operator is related to the pion,
the Goldstone boson of chiral symmetry breaking.  This state becomes massless for
vanishing $\mpcac$ and is the lowest state in the spectrum at small
enough fermion mass. In several works related to theories different
from QCD one adopts the convention to call pions the Goldstone bosons
related to chiral symmetry breaking and define parity in such a way
that these states are pseudoscalar.  In investigations of supersymmetric
Yang--Mills theory, for example, the chiral symmetry breaking pattern  
is defined in a partially quenched setup~\cite{Munster:2014cja} and
the light meson is called adjoint pion. Under the Lorentz group, the corresponding creation
operator can be made to transform as a pseudoscalar if one deviates from the
definition of the parity \eqref{eq:parity}, which is the same as in
QCD. A consequence of that definition is that Majorana
flavours are mixed by a parity transformation:
\begin{align}
 \xi_{+}(t,\vec{x}) &\mapsto i\gamma_0\xi_{-}(t,-\vec{x})\; .
\end{align}
If one uses instead 
\begin{align}
\psi(t,\vec{x}) &\mapsto i\gamma_0\gamma_5 \psi(t,-\vec{x}) \;, \\
\pb(t,\vec{x}) &\mapsto -i\pb(t,-\vec{x})\gamma_0\gamma_5\;,
\end{align}
the two Majorana flavours do not mix. With this choice, the parity quantum
numbers of what in our conventions are the baryonic states are
interchanged (pseudoscalar becomes scalar and vice versa). 
Hence what is called in other investigations a pseudoscalar meson and
the (adjoint) pion, in our study is the
scalar baryon. While the convention used elsewhere might seem more
natural from the point of view of associating states to their QCD
equivalent, ours treats on equal footing
the left and right component of the Dirac spinor, as it happens for
the standard definition of the parity in QCD. Ultimately, were
the theory studied in this work to be found in nature, the
interaction of its particles with the standard model particles would
provide a natural way of fixing the arbitrariness in the definition of
the parity.\\
\section{Correlation functions of fermion bilinears}
\label{app:c}
Our nomenclature for the contraction of correlation functions borrows
from that of QCD for the sake of familiarity. Consider a correlation
function for a theory with $N_{\textnormal{f}}\ne1$, 
\begin{equation}
	I(x,y) = \frac{1}{Z} \int DU D\pb D\psi \,\bcontraction{}{\pb}{_a\overline{\Gamma}}{\psi}\contraction[2ex]{}{\pb}{_a\Gb\psi_b(x)\,\pb_a\Gamma}{\psi} \pb_a \Gb \contraction{}{\psi}{_b(x),}{\pb}\psi_b(x)\,\bcontraction{}{\pb}{_a\Gamma}{\psi}\pb_a\Gamma\psi_b(y) e^{-S}\;,
\end{equation}
where $a$, $b$ are flavour indices, to be contracted with Wick's
theorem, and $Z$ is the path integral. In the case $a\ne b$, then only
the upper contraction gives a non-zero contribution, which then
becomes a term of the form $-\tra \Gb D^{-1}(x; y) \Gamma D^{-1}(y;
x)$, which we refer to as both the triplet (this name being inspired
by the isospin symmetry of QCD) and the connected contribution (where
the terminology refers to the fact that in terms of purely fermionic
lines the corresponding diagram is connected). If $a=b$, however, both
contractions can give non-zero contributions, resulting in a linear
combination of the previous term and one of the form  $\tra \Gb
D^{-1}(x;x) \tra\Gamma D^{-1} (y;y)$. The latter term we refer to as the
disconnected contribution, while the linear combination we call the
singlet. The reason why triplet correlation functions may give
physically meaningful states in what is a 1-flavour theory (where one
would na\"ively assume $a=b$, so only singlets are valid) is
discussed in the main body of the text. 

\bibliography{references}

\end{document}

%% file: tables.tex
 \begin{table}
	 \caption{The lattices considered in this study. Here
           $N_{\textnormal{conf}}$ indicates the number of
           thermalised configurations used in the averages, $a m$ is the bare fermion
           mass in units of the lattice spacing $a$
           and the first column is a reference name for the set. Also indicated for each set  is
           the lattice volume. \label{tab:lattices}}
	 \begin{ruledtabular}
		 \begin{tabular}{cccc}
			 Name & Volume & $-am$ & $N_{\textnormal{conf}}$ \\
			 \hline
			A1	&	$16\times8^3$	&	$1.475$	&	1500	\\
			A2	&	$16\times8^3$	&	$1.500$	&	1500	\\
			A3	&	$16\times8^3$	&	$1.510$	&	1500	\\
			A4	&	$16\times8^3$	&	$1.510$	&	4000	\\
			\hline
			B1	&	$24\times{12}^3$	&	$1.475$	&	1500	\\
			B2	&	$24\times{12}^3$	&	$1.500$	&	1500	\\
			B3	&	$24\times{12}^3$	&	$1.510$	&	4000	\\
			\hline
			C1	&	$32\times{16}^3$	&	$1.475$	&	1500	\\
			C2	&	$32\times{16}^3$	&	$1.490$	&	1500	\\
			C3	&	$32\times{16}^3$	&	$1.510$	&	1500	\\
			C4	&	$32\times{16}^3$	&	$1.510$	&	4000	\\
			C5	&	$32\times{16}^3$	&	$1.514$	&	1500	\\
			C6	&	$32\times{16}^3$	&	$1.519$	&	1500	\\
			C7	&	$32\times{16}^3$	&	$1.523$	&	1500	\\
			\hline
			D1	&	$48\times{24}^3$	&	$1.510$	&	1534	\\
			D2	&	$48\times{24}^3$	&	$1.523$	&	2168	\\
			 
		 \end{tabular}
	 \end{ruledtabular}
 \end{table}

\begin{table}[htbp]
	\centering
	\caption{Glueball masses and string tension}
	\label{tab:glue}
	\begin{ruledtabular}
		\begin{tabular}{c | c c c}
			Name	&	$a \sqrt{\sigma}$	&	$am_{0^{+}}$	&	$am_{2^{+}}$ \\
			\hline
			A1	&	0.424(13)	&	$0.8422\pm0.0968$	&	$1.3148\pm0.2305$ \\
			A2	&	0.335(10)	&	$0.7320\pm0.0885$	&	$1.4678\pm0.2176$ \\
			A3	&	0.299(12)	&	$0.5690\pm0.0585$	&	$1.6921\pm0.3196$ \\
			A4	&	--- & $0.5873\pm0.0553$	&	--- \\
			\hline
			B1	&	0.378(19)	&	$0.9582\pm0.1174$	&	$1.8059\pm0.3643$ \\
			B2	&	---	&	$0.7296\pm0.1092$	&	--- \\
			B3	&	0.322(10)	&	$0.5284 \pm 0.1494$	&	--- \\
			\hline
			C1	&	0.436(60)	&	$0.9654\pm0.1057$	&	$1.7461\pm0.3526$ \\
			C2	&	0.379(44)	&	$0.8265\pm0.0644$	&	$1.9130\pm0.5004$ \\
			C3	&	0.318(11)	&	$0.5985\pm0.0573$	&	$1.6285\pm0.3079$ \\
			C4	&	---	&	$0.5901\pm0.0438$	&	--- \\
			C5	&	0.322(13)	&	$0.5530\pm0.0415$	&	$1.5834\pm0.2263$ \\
			C6	&	0.2859(75)	&	$0.3689\pm0.0437$	&	$1.9897\pm0.2589$ \\
			C7	&	0.2368(84)	&	$0.3146\pm0.0278$	&	$1.0188\pm0.0977$ \\
			\hline
			D1	&	---	&	$0.4609\pm0.0553$	&	--- \\
			D2	&	0.2354(56)	&	$0.3595\pm0.0219$	&	$1.3387\pm0.1104$ \\
		\end{tabular}
	\end{ruledtabular}
\end{table}

\begin{table}[htbp]
	\centering
	\caption{PCAC and baryon masses (triplet channels).}
	\label{tab:mesons}
	 \begin{ruledtabular}
		\begin{tabular}{ c | c c c c}
			Name	&	$am_{\mathrm{PCAC}}$	&	$am_{\textnormal{pseudoscalar}}$ & $am_{\textnormal{scalar}}$ & $am_{\textnormal{vector}}$\\
			\hline
			A1    &	0.1486(17)		&	--- 	&	0.9704(58)	&	---\\
			A2	&	0.1108(20)	&	--- 	&	0.8432(81)	&	---\\
			A3	&	0.0906(27)	&	--- 	&	0.763(12)	&	---\\
			A4	&	0.0872(22)	&	--- 	&	0.747(10)	&	---\\
			\hline
			B1	&	0.1493(29)	&	--- &		0.9733(23)	& 	2.297(59)\\
			B2	&	0.1113(8)	&	1.969(39) &	0.8449(31)	&	2.062(41)\\
			B3	&	0.0911(7) &	1.635(45) &	0.7644(30)	&	--- \\
			\hline
			C1	&	0.1490(3)	&	---	 &	0.9723(12)	&	---\\
			C2	&	0.1278(3)	&	--- 	&	0.9035(16)	&	---\\
			C3	&	0.0911(3)	&	---	&	0.7646(15)	&	---\\
			C4	&	0.0905(5)	&	1.594(58) &	0.7645(17)	&	---\\
			C5	&	0.0829(6)	&	1.712(31) &	0.7288(29)	&	1.702(70)\\
			C6	&	0.0659(9)	&	1.518(42) & 	0.6473(44)	&	--- \\
			C7	&	0.0484(5)	&	--- 	&	0.5480(36)	&	---\\
			\hline
			D1	&	0.0913(2)	&	---	&	0.7651(11)	&	---\\
			D2	&	0.0472(3)	&	1.282(62) &	0.5412(25)	&	--- \\
		\end{tabular}
	\end{ruledtabular}
\end{table}

\begin{table}[htbp]
	\centering
	\caption{Meson and \spinhalf state masses (singlet channels).}
	\label{tab:singlet}
	\begin{ruledtabular}
		\begin{tabular}{c | c c c c c}
			Name	&	$a m_{1/2}$ & $a m_{\textnormal{scalar}}$ & $a m_{\textnormal{pseudoscalar}} $ &   $a m_{\textnormal{axial}}$& $a m_{\textnormal{vector}}$\\
			\hline 
			B1&1.707(20)   &  0.64(20) & 0.9859(91) & 1.1575(33) & --- \\
			B2&1.579(54)  & --- & 0.839(45) &  1.0496(61) & 1.91(10) \\
			B3&1.3519(94)  & --- & 0.778(49) &  --- & --- \\ 
			\hline
			C4&1.325(24) & 0.65(21) & 0.751(19) &  --- & --- \\
			C5&1.240(57) & 0.403(80) & 0.750(17) & 0.938(10) & --- \\
			C6&1.135(48)  & 0.296(96) & 0.661(82) & 0.8551(75) & 2.190(52) \\ 
			\hline
			D1 & --- & --- & --- & 0.9761(53) & --- \\
			D2&1.013(17) & 0.28(12) & 0.551(26) & 0.7412(90) & --- \\
		\end{tabular}
	\end{ruledtabular}
\end{table}